\documentclass[12pt]{iopart}

\usepackage{iopams}
\usepackage{amssymb}
\usepackage{amsthm}
\usepackage{graphicx}
\usepackage{hyperref}
\usepackage{dcolumn}
\usepackage{bm}
\usepackage{multirow}
\usepackage{comment}
\usepackage{array,booktabs}
\usepackage{braket}
\usepackage{cite}
\usepackage{xcolor}
\usepackage{color}
\usepackage{url}
\usepackage[normalem]{ulem}

\hypersetup{
    colorlinks=true,
    citecolor=blue,
    linkcolor=blue
}

\begin{document}

\title[Dynamical Regimes of Discrete Diffusion Models]{Dynamical Regimes of Discrete Diffusion Models}

\author{Tomoei Takahashi $^{1}$, Takashi Takahashi $^{2,4}$, \\and Yoshiyuki Kabashima $^{2,3}$}

\address{$^1$ Department of Mathematical and Systems Engineering, Shizuoka University, 3-5-1 Johoku, Chuo-ku, Hamamatsu, 432-8561, Japan.}
\address{$^2$ Institute for Physics of Intelligence, The University of Tokyo, 7-3-1 Hongo, 113-0033, Japan.}
\address{$^3$ Trans-Scale Quantum Science Institute, The University of Tokyo, 7-3-1 Hongo, 113-0033, Japan.}
\address{$^4$ RIKEN Center for Advanced Intelligence Project (AIP), RIKEN, Chuo, Tokyo 103-0027, Japan}
\ead{takahashi.tomoei@shizuoka.ac.jp}
\vspace{10pt}
\begin{indented}
\item[]
\end{indented}

\begin{abstract}

Diffusion models generate high-dimensional data such as images by learning a process that gradually removes noise from corrupted data. 
Recent studies have shown that the backward dynamics of diffusion models exhibit two characteristic transitions: the speciation transition, at which generated samples begin to capture the global structure of the training data, and the collapse transition, at which the generation dynamics starts committing to individual training samples.
While these transitions have been theoretically analyzed for continuous data, the same theoretical criteria have not been applied for discrete diffusion models, which are diffusion models for discrete data with important applications such as language and graph data. It is nontrivial whether the theoretical framework that successfully describes these transitions for continuous data remains valid for discrete variables, whose state space is not continuously distributed.
In this work, we propose a simple effective model for discrete diffusion models trained on two-class Ising variable data with a general mixture ratio and analyze its backward dynamics using methods from statistical mechanics. We show that, as in the previous study on continuous data, the speciation transition can be determined through a second-order phase transition analysis using high-temperature expansion, while the collapse transition corresponds to a condensation transition described by the Random Energy Model. An analytical expression for the speciation time is obtained, and we show that its scaling becomes consistent with that of the continuous case when the noise increases with time as in practical diffusion models.
These theoretical predictions are confirmed by numerical simulations and experiments with trained discrete diffusion models on real datasets.
These results suggest that the original theoretical framework for continuous data remain valid for discrete data, and may provide a useful starting point for the statistical-mechanics analysis of generative diffusion dynamics for discrete variables in more realistic settings.

\end{abstract}

\vspace{2pc}
\noindent{\it Keywords}: Discrete diffusion models, Speciation, Collapse
%\maketitle

\tableofcontents
\addtocontents{toc}{\protect\thispagestyle{empty}}
\pagestyle{empty}

\section{Introduction}
\label{sec:intro}

Diffusion models have recently achieved remarkable success across a wide range of applications, including image and video generation, attaining state-of-the-art generative performance \cite{sohl2015deep, ho2020denoising, song2019generative, yang2023diffusion}. Diffusion models consist of two stochastic processes: a forward process and a backward process. In the backward process, the model generates samples by progressively removing noise.

One of the major fundamental mysteries of diffusion models is the origin of their generalization ability. This refers to the capability to generate data that are very similar to the training data but do not appear in the training set itself.
A straightforward approach to clarifying the origin of the generalization ability of diffusion models is to theoretically analyze the dynamics of their learning process \cite{bonnaire2025diffusion, kamb2025an, cui2024analysis}.

However, in this paper, we do not directly address this problem of generalization. Instead, as a first step toward tackling this broader question, we follow the pioneering work \cite{biroli2024dynamical} and focus on the dynamical properties of the backward process of diffusion models for discrete data under the assumption of ideal learning. 
A major advantage of this approach is that it allows us to focus on the intrinsic dynamics of generation under ideal learning, without entangling it with the difficult question—often raised in the context of diffusion models—of whether generalization simply results from imperfections in the learning process.

For continuous Gaussian data, the trajectories of the reverse process were analyzed in \cite{biroli2024dynamical}, building on a rigorous framework for high-dimensional generative diffusion dynamics developed in \cite{biroli2023generative}, and three distinct dynamical regimes were identified. These are: (I) a Brownian-like regime in which trajectories wander randomly; (II) a regime in which trajectories capture the global structure of the training data and dynamically converge toward a specific class; and (III) a regime in which trajectories further converge toward a particular training sample within that class. The transition from regime I to regime II is termed speciation, while the transition from regime II to regime III is termed collapse.

Building on this line of research, several statistical-mechanics analyses of the speciation and collapse transitions have been developed, mainly for continuous-valued data. Speciation has been characterized in terms of the free entropy difference between classes \cite{achilli2026theory}. The collapse transition has been interpreted as a condensation transition in the Random Energy Model \cite{achilli2025memorization} and as a collapse of the tangent subspace of the data manifold \cite{achilli2024losing}. In addition, the three dynamical regimes described above have also been characterized from the geometric dynamics of the data structure \cite{ventura2025manifolds, ventura2026emergence}.

However, aside from the analysis of the one-dimensional Ising chain in \cite{achilli2026theory}, these studies have primarily focused on continuous data, either generic continuous distributions or data satisfying the manifold hypothesis. For discrete data with applications as important as those of continuous data \cite{li2022diffusion, lin2023text, ye2024diffusion, austin2021structured, hoogeboom2022autoregressive}, the manifold hypothesis assumed for continuous data does not necessarily hold. Therefore, the geometric approach based on the structure of the generation process described above cannot be directly applied. In this work, we thus examine whether the original criteria for the dynamical phase boundaries remain valid for discrete data. To this end, we directly apply the phase boundary criteria proposed in \cite{biroli2024dynamical} to the generation dynamics of discrete diffusion models. Whether the phase boundary criteria derived for continuous data remain applicable to discrete data is a nontrivial question and constitutes an important issue for understanding the dynamical properties of discrete diffusion models.

In the following, we first describe discrete diffusion models in Sec. \ref{sec:sec2}.
We then propose in Sec. \ref{sec:sec3} an effective model for the theoretical analysis of discrete diffusion models using $\pm 1$ Ising-variable data with a general two-class mixture ratio $\eta \in [0,1]$. In Sec. \ref{sec:sec4}, we present the theoretical details for determining the speciation time (Sec. \ref{subsec:sec4_1}) and the collapse time (Sec. \ref{subsec:sec4_2}). In Sec. \ref{sec:sec5}, we report the results of numerical experiments validating the theoretical predictions for the speciation time in both class-balanced and class-imbalanced cases (Sec. \ref{subsec:sec5_1}), as well as similar validation results for the collapse time (Sec. \ref{subsec:sec5_2}). Finally, in Sec. \ref{sec:sec6} we present experimental results on real datasets for the respective transition points.

\section{Discrete diffusion models}
\label{sec:sec2}

Discrete diffusion models, similarly to standard diffusion models, are latent-variable models consisting of a forward process and a backward process, both formulated as Markov processes.
Given a training sample $\bm x_{0}$, the forward process of a discrete diffusion model generates a sequence of noise-perturbed variables $\bm x_{1}, \bm x_{2}, \cdots, \bm x_{T}$, where $\bm x_{t} = (x_{t1}, x_{t2}, \cdots, x_{tN})^{\top}$ denotes an $N$-dimensional data vector at time $t$. Here and throughout the paper, $(\cdots)^{\top}$ denotes the transpose of a vector or matrix.
Each component $x_{ti}$ of $\bm x_{t}$, for $i = 1,2,\ldots,N$, takes one of $K$ categorical values. That is, $x_{ti} \in \{1,\ldots,K\}$. The index $i$ represents the position within a single data point, as in the continuous case. Hence, each $i = 1,2, \cdots$ denotes the index of each pixel or token within a single data point.

In the forward process, the addition of noise is represented by stochastic flips between categories. In discrete diffusion models, noise is given by the flip probability of the state of each variable.
The transition probabilities between categories that govern this process are described by the $K \times K$ transition probability matrix $\bm Q_{t}$ in which the $(a,b)$-th element is given by $\bm [\bm Q_{t}]_{ab} = q(x_{ti} = a | x_{t-1,i} = b)$, where $a,b = 1,\ldots,K$. In this section, following the standard description of discrete diffusion models, each element of data vector at time $t$, $x_{ti}$ for $t = 0,1,\cdots, T$, and $i = 1,2,\cdots,N$, is represented by a one-hot column vector $\bm z_{ti}$. That is, when $x_{ti} = k$, the vector $\bm z_{ti}$ satisfies $z_{ti,k} = 1$ and $z_{ti,j} = 0$ for all $j \ne k$. Throughout this work, we use both the index-based representation and the one-hot representation depending on the context.

The probability distribution of the forward process is as follows:
\begin{eqnarray}
    q(\bm z_{ti} | \bm z_{t-1,i}) = \mathrm{Cat}(\bm z_{ti} | \bm p = \bm Q_{t} \bm z_{t-1,i}),
\end{eqnarray}
where
\begin{eqnarray}
\mathrm{Cat}(\bm x | \bm p) = \prod_{k = 1}^{K} p_{k}^{x_{k}}\label{eq def categorical dist},\\
\sum_{k = 1}^{K} p_{k} = 1 .
\end{eqnarray}
Eq. (\ref{eq def categorical dist}) is the categorical distribution parameterized by a $K$-dimensional vector $\bm p$ whose components represent the probabilities of each category. 
The categorical distribution is the multinomial generalization of the Bernoulli distribution.
All variables (e.g., pixels in images or tokens in language data) evolve independently in time according to the transition probability $q(\bm z_{ti} | \bm z_{t-1,i})$, irrespective of the index $i$. The transition matrix $\bm Q_{t}$ is shared across all variables.
Furthermore, the transition probability from time $0$ to time $t$, denoted by $q(\bm x_{t} | \bm x_{0})$ (where the index $i$ is omitted since the transition probability is the same for all $i = 1,2,\cdots,N$), is given by
\begin{eqnarray}
    q(\bm z_{t} | \bm z_{0}) = \mathrm{Cat}(\bm z_{t} | \bm p = \overline{\bm {\bm Q_{t}}}\bm z_{0} ),
\end{eqnarray}
where $\overline{{\bm Q_{t}}}$ is defined by
\begin{eqnarray}
    \overline{{\bm Q_t}} = \bm Q_{t}\bm Q_{t-1} \cdots \bm Q_{1}.
\end{eqnarray}
The choice of the transition probability matrix is an important issue in discrete diffusion models.
The simplest choice is the uniform transition, in which the transition probabilities between different categories are taken to be constant \cite{hoogeboom2022autoregressive}.
In the uniform-type transition, the transition matrix is given by
\begin{eqnarray}
\bm Q_{t} = (1 - \beta_{t}) \bm I + \frac{\beta_{t}}{K} \bm 1,
\end{eqnarray}
where $\beta_{t} \in [0,1]$ is the parameter controlling the noise level at time $t$ and represents the probability of transitioning to a different category, and $\bm I$ and $\bm 1$ denote the $K \times K$ identity matrix and the $K \times K$ matrix with all entries equal to one, respectively.

Here we provide a description of the learning procedure for discrete diffusion models. However, as mentioned above, the problem of learning is absent in the present study. We therefore only present a brief sketch. 
In discrete diffusion models, the state space is discrete, and therefore the score function used in diffusion models for continuous data cannot be computed directly. Instead, the backward transition probability $p_{\Theta}(\bm z_{t-1} | \bm z_{t})$ is modeled as a categorical distribution parameterized by a neural network, whose parameters are learned by minimizing the cross-entropy between the predicted distribution and the true one-hot state $\bm z_{t-1}$ produced by the forward process. A more detailed explanation is provided in Appendix \ref{app:appC}.

In the following analysis of discrete diffusion models, we consider a setting in which the probability distribution of the data, $q_{0}$, is specified explicitly, so that the marginal distribution $q_{t}$ is known.
In this case, the learning problem is eliminated, and the remaining task reduces to sampling accurately from this distribution.
Indeed, for the effective discrete diffusion model proposed below, we construct an efficient sampling method that is exact in the limit $N \rightarrow \infty$.

\section{Effective model of discrete diffusion models}
\label{sec:sec3}

We propose a simple ``model" of discrete diffusion models. The discrete data are the set of the $N$-Ising spin system: $\bm x_{0} = (x_{01}, x_{02}, \cdots, x_{0N}), x_{0i} \in \{-1, +1\} (i = 1,2,\cdots, N)$. We denote $\bm x_{t} \in \{-1,+1\}^{N}$ as the noise-perturbed data at time $t$ in the forward and backward process,
where the parameter $m \in [0,1]$ is the probability that any $x_{0i}$ in the first-term class is $+1$, and in the second-term class is $-1$. The parameter $\eta \in [0,1]$ is the control parameter of the ratio of each class.
We assume a two-component mixture distribution for the data at $t=0$ as
\begin{equation}
	P_0(\bm{x}_0) = \eta P^+(\bm{x}_0) + (1-\eta)P^-(\bm{x}_0), \quad \eta \in [0,1]\label{def P0}
\end{equation}
where
\begin{equation}
	P^+(\bm{x}_0) = \prod_{i=1}^N \frac{1+mx_{0i}}{2},
	\quad 
	P^-(\bm{x}_0) = \prod_{i=1}^N \frac{1-mx_{0i}}{2}\label{eq def P_plus and P_minus}.
\end{equation}
In each component $P^\pm$, the variables $x_{0i}$ are independent and identically distributed with mean $\pm m$.
The probability distribution of the forward process is as follows: 
\begin{eqnarray}
    P(\bm x_{t} | \bm x_{t-1}) = \prod_{i = 1}^{N}\frac{1 + \theta x_{ti}x_{t-1,i}}{2},
\end{eqnarray}
%where $\theta$ relates to the noise schedule. For every time step, the probability of spin flip is defined as $\mathrm{Pr}(flip) = (1-\theta)/2$. The definition of parameter $\theta$ is as follows: $\theta = 1 - \gamma/N$. Thus, $\gamma$ represents the number of spins that will be flipped at each time. We assume $\gamma = \mathcal{O}(1)$. Therefore, $\theta$ represents the fraction of spins that remain fixed without flipping.
where $\theta$ relates to the noise schedule. For every time step, the probability of spin flip is defined as $\mathrm{Pr}(\mathrm{flip}) = (1-\theta)/2$. The definition of parameter $\theta$ is as follows: $\theta = 1 - \beta$, where $\beta$ is set to $0 < \beta \ll 1$. The parameter $\theta$ represents the fraction of spins that remain fixed without flipping.

This setting is equivalent to the formulation described in Sec. \ref{sec:sec2} with $K = 2$, where the transition probability matrix $\bm Q_{t}$ is specified as follows.
\begin{eqnarray}
\bm Q_{t} =
\left(
\begin{array}{cc}
\frac{1+\theta}{2} & \frac{1-\theta}{2}\\
\frac{1-\theta}{2} & \frac{1+\theta}{2}
\end{array}
\right)
\end{eqnarray}
With this choice, the product of the transition matrices up to time $t$ can be written as
\begin{eqnarray}
\overline{\bm Q_{t}} =
\left(
\begin{array}{cc}
\frac{1+\theta^{t}}{2} & \frac{1-\theta^{t}}{2}\\
\frac{1-\theta^{t}}{2} & \frac{1+\theta^{t}}{2}
\end{array}
\right),
\end{eqnarray}
Thus, the transition probability from time $0$ to time $t$, $P(\bm x_{t} | \bm x_{0})$, can be written in a particularly simple form as follows.
\begin{eqnarray}
    P(\bm x_{t}|\bm x_{0}) = \prod_{i = 1}^{N}\frac{1 + \theta^{t} x_{ti} x_{0i}}{2}\label{eq forward 0tot}.
\end{eqnarray}
Because we already know the data distribution $P(\bm x_{0})$, we can obtain the marginal distribution of each time step in the forward process $P(\bm x_{t})$ by the marginalization: $\sum_{\bm x_{0}} P(\bm x_{t} | \bm x_{0}) P_{0}(\bm x_{0}) = P_{t}(\bm x_{t})$. The formula becomes the following.
\begin{eqnarray}
    P_{t}(\bm x_{t}) &= \eta\prod_{i = 1}^{N}\frac{1 + \theta^{t} m x_{ti}}{2} + (1-\eta) \prod_{i = 1}^{N}\frac{1 - \theta^{t} m x_{ti}}{2}\label{eq def Pt}.
\end{eqnarray}
Since the marginal distribution $P_{t}(\bm x_{t})$ has been obtained, we can derive the probability distribution of reverse process $P(\bm x_{t-1} | \bm x_{t})$ through the Bayes's theorem: $P(\bm x_{t-1} | \bm x_{t}) = \frac{P(\bm x_{t} | \bm x_{t-1}) P_{t-1}(\bm x_{t-1})}{P_{t}(\bm x_{t})}$. However, the direct sampling from the backward process obtained by substituting to the Bayes' theorem with the previously derived probability distributions is still difficult. In this study, we propose a method to efficiently obtain accurate data samples with no approximation in the limit $N \rightarrow \infty$. \footnote{This method is based on the idea of Koki Okajima.} (for details, see Appendix \ref{app:appA}).

\section{Theoretical analysis}\label{sec:sec4}
\subsection{The speciation time}\label{subsec:sec4_1}

Speciation is the moment when a clear macroscopic direction can be discerned from trajectories that otherwise move randomly in a Brownian‑like motion. This is like a thermodynamic phase transitions, for instance, the case a ferromagnetic system develops a non-zero magnetization. This type of transition can be studied by a perturbative expansion\cite{Chaikin_Lubensky_1995}.

% However, in the form of $P_{t}(\bm x_{t})$ given by Eq. (\ref{eq def Pt}), the distribution consists of two components and therefore cannot be treated as a Boltzmann-distribution-like form, making such a statistical mechanics analysis difficult. Then we think $P(\bm x_{t} | \bm x_{0})$, and perform the marginalization $P_{t}(\bm x_{t}) = \sum_{\bm x_{0}} P(\bm x_{t} | \bm x_{0}) P_{0}(\bm x_{0})$. This is because, as shown below, $P(\bm x_{t} | \bm x_{0})$ can be transformed into a form written using an exponential function and thus rewritten as a Boltzmann-distribution-like form: 
However, $P_t(\bm x_t)$ is a mixture distribution, and this makes an analysis difficult. To overcome this difficulty, we use the representation
\begin{equation}
P_t(\bm x_t)=\sum_{\bm x_0} P(\bm x_t | \bm x_0)P_0(\bm x_0).
\end{equation}
Although this marginalization is straightforward, we can use a high-temperature expansion of $P(\bm{x}_t| \bm{x}_0)$ in $F_t:=\tanh^{-1} \theta^{t}$, by assuming that the speciation transition occurs at $t\gg1$. This yields an effective Ising-like description for $P_t(\bm{x}_t)$, which is much easier to analyze.

Specifically, we use the identity
\begin{eqnarray}
\frac{1 + \tanh(F )S}{2} = \frac{e^{F S}}{2 \cosh F },
\qquad F\in\mathbb{R}, \quad S\in\{-1,1\}
\label{eq def binom to exp},
\end{eqnarray}
and rewrite Eq. (\ref{eq forward 0tot}) as
\begin{eqnarray}
P(\bm x_{t} | \bm x_{0}) = \frac{e^{F_{t} \sum_{i = 1}^{N}x_{ti}x_{0i}}}{[2\cosh F_{t} ]^{N}}\label{eq Pt_0 exp form}.
\end{eqnarray}

As discussed above, speciation is the phenomenon in which data trajectories leave the Brownian-like regime and begin to capture the global features of the data. Equivalently, if we consider the forward-time direction, the occurrence of speciation indicates that the signal from the training data has already become very weak at the corresponding time. Therefore, it is reasonable to assume that $\theta^t \ll 1$ around the speciation time, since $\theta^t$ can be regarded as the strength of the signal from the initial data at time $t$. Consequently, the “inverse temperature” $F_t$ satisfies $F_t=\tanh^{-1}\theta^t=\theta^t+o(\theta^t)\ll1$, and a high-temperature expansion of Eq. (\ref{eq Pt_0 exp form}) in powers of $F_t$ can be performed around the speciation time.

The same reasoning applies to time-dependent noise schedules by replacing $\theta^t$ with the cumulative signal $\Theta_t:=\prod_{s=1}^{t}(1-\beta_s)$. Near the speciation time, $\Theta_t\ll1$, and hence $F_t=\tanh^{-1}\Theta_t=\Theta_t+o(\Theta_t)\ll1$. Therefore, the same high-temperature expansion remains applicable.

The high-temperature expansion is
\begin{eqnarray}
    \hspace{-20mm} P(\bm x_{t} | \bm x_{0}) = \left( 2\cosh F_{t}\right)^{-N} \left( 1 + F_{t}\sum_{i = 1}^{N} x_{ti}x_{0i} + \frac{1}{2} F_{t}^{2} \sum_{i,j = 1}^{N} x_{ti}x_{tj}x_{0i}x_{0j} + o(F_{t}^{2}) \right).
\end{eqnarray}
Let $\langle \cdot \rangle$ denotes the expectation under $P_{0}(\bm x_{0})$, we obtain
\begin{eqnarray}
\hspace{-20mm}\log P_{t}(\bm x_{t})
&=&
\log \sum_{\bm x_{0}} P_{0}(\bm x_{0}) P(\bm x_{t} | \bm x_{0}) \\
&\approx&
- N \log \left( 2 \cosh F_{t} \right)
+ \log \left(
1
+ F_{t} \sum_{i=1}^{N} x_{ti} \langle x_{0i} \rangle
+ \frac{1}{2} F_{t}^{2} \sum_{i,j=1}^{N} x_{ti} x_{tj}
\langle x_{0i} x_{0j} \rangle
\right) \nonumber\\
&\approx&
- N \log \left( 2 \cosh F_{t} \right)
+ F_{t} \sum_{i=1}^{N} x_{ti} \langle x_{0i} \rangle
\nonumber\\
&&
+ \frac{1}{2} F_{t}^{2}
\sum_{i,j=1}^{N} x_{ti} x_{tj}
[
\langle x_{0i} x_{0j} \rangle
-
\langle x_{0i} \rangle \langle x_{0j} \rangle
]\\
&=&
- N \log \left( 2 \cosh F_{t} \right)
+ F_{t} \sum_{i=1}^{N} x_{ti} \langle x_{0i} \rangle +\frac{1}{2}F_{t}^{2}\sum_{i = 1}^{N}\left(1-\langle x_{0i} \rangle^{2} \right) 
\nonumber\\
&&
+
\frac{1}{2}F_{t}^{2} \bm x_{t}^{\top} \bm J \bm x_{t}\label{eq landau expansion of logPt}.
\end{eqnarray}
Here, the $(i,j)$-th element of matrix $\bm J$ is given by $J_{ij} = (1- \delta_{ij})[\langle x_{0i}x_{0j} \rangle-\langle x_{0i} \rangle \langle x_{0j} \rangle]$. We used that the diagonal terms of the second-order term become constant (the third term of Eq. (\ref{eq landau expansion of logPt})).

The above results show that $P_{t}(\bm x_{t})$ can be regarded as the Boltzmann distribution with the inverse temperature $F_{t}$ and the Hamiltonian
\begin{eqnarray}
    H(\bm x_{t}) = -\frac{1}{2} F_{t}\sum_{i \neq j}J_{ij}x_{ti}x_{tj} - \sum_{i =1}^{N} x_{ti} \langle x_{0i}\rangle.
\end{eqnarray}
We perform the mean-field approach to this Hamiltonian. The self-consistent equation is
\begin{eqnarray}
    m_{ti} = \tanh \left( F_{t}^{2} \sum_{j \neq i} J_{ij} m_{tj} + F_{t}\langle x_{0i}\rangle \right)\label{eq TAP},
\end{eqnarray}
where $m_{ti}$ is the thermal average of $x_{ti}$, namely the expectation under $P_{t}(\bm x_{t})$ as the Boltzmann distribution. When the system is asymmetric under spin reversal, namely when $\eta \neq 0.5$ so that $\langle x_{0i} \rangle \neq 0$, a nonzero external field is present. This may allow for a first-order transition. However, in the present case, since $F_{t} \ll 1$, the magnitude of the external field is sufficiently small, and we assume that it can be neglected. Because $0 < F_{t} \ll 1$, we get
\begin{eqnarray}
    m_{ti} = F_{t}^{2} \sum_{j \neq i} J_{ij}m_{tj} + F_{t} \langle x_{0i} \rangle + o(F_{t}^{3})\label{eq TAP expanded}.
\end{eqnarray}
Defining $\bm m_{t} = (m_{t1}, m_{t2}, \cdots, m_{tN})^{\top}$ and $\bm x_{0} = \left( x_{01}, x_{02}, \cdots x_{0N}\right)^{\top}$, we can obtain the following from Eq. (\ref{eq TAP expanded})
\begin{eqnarray}
    \bm m_{t} \approx \left( \bm I - F_{t}^{2} \bm J\right)^{-1} F_{t} \langle \bm x_{0} \rangle,
\end{eqnarray}
where $\bm I$ denotes $N \times N$ identity matrix. 
The speciation transition, here namely the second-order transition occurs at the point at which $\bm m_{t}$ diverges. Hence the speciation time $t_{S}$ satisfies the following relation
\begin{eqnarray}
    1 = F_{t_{S}}^{2} \Lambda,\label{eq def noise schedule free ts criterion}
\end{eqnarray}
where we denoted by $\Lambda$ the maximum eigenvalue of $\bm J$.
Eq. (\ref{eq def noise schedule free ts criterion}) gives a general, noise-schedule-independent criterion for the speciation time $t_{S}$.

Since $F_{t} = \theta^{t} + o(\theta^{t})$, it follows that $F_{t}^{2} \approx \theta^{2t} = (1-\beta)^{2t}$ in the same manner as the above approximation. Then, we get
\begin{eqnarray}
    t_{S} = -\frac{\log \Lambda}{2\log(1-\beta)}.
\end{eqnarray}
 Since $0 < \beta \ll 1$, $\log (1 - \beta) \approx -\beta$. Hence, we finally obtain the simple analytic expression of $t_{S}$ as follows:
\begin{eqnarray}
    t_{S} = \frac{1}{2\beta}\log \Lambda \label{eq ts}.
\end{eqnarray}
Since $\Lambda$ is of order $N$, $t_{S} \gg 1$ holds when $N \gg 1$. This ensures the self-consistency of the assumptions introduced above, and justifies the subsequent analysis in the regime $t_{S}\gg1$.

We now discuss the correspondence with the result for the speciation time in the continuous case derived in \cite{biroli2024dynamical}. In \cite{biroli2024dynamical}, the forward process is a Gaussian distribution whose mean and variance are $e^{-t}\bm{x}_0$ and $1-e^{-2t}$, respectively. In the theory of recent standard diffusion models, Denoising Diffusion Probabilistic Models (DDPMs) \cite{ho2020denoising}, the mean value of the probability distribution of the forward process is the square root of the cumulative signal factor at each time $t$. Hence, the cumulative signal factor becomes $\Theta_t=\prod_{s=1}^{t}(1-\beta_s)=e^{-2t}$. Since we consider the regime in which the cumulative signal is small as mentioned above, the corresponding inverse temperature satisfies $F_t\approx \Theta_t$. Therefore, in the general criterion Eq. (\ref{eq def noise schedule free ts criterion}), we may replace $F_t$ by $e^{-2t}$. This gives
\begin{eqnarray}
t_S=\frac{1}{4}\log\Lambda \label{eq ts correspond to continuous}.
\end{eqnarray}
In \cite{biroli2024dynamical}, the speciation time is obtained as $t_S=\frac{1}{2}\log\Lambda$. Thus, our result reproduces the same leading logarithmic scaling with respect to $\Lambda$.

Moreover, in \cite{achilli2026theory}, the speciation time for both continuous and discrete variables is defined through a general criterion: it is the time at which the average free entropy difference between competing classes becomes comparable to the typical fluctuation of that difference. Their framework covers the case of an arbitrary number of classes, including situations in which the classes cannot be distinguished by their first moments. As a result, they showed that, for the one-dimensional Ising chain, the speciation time behaves as $t_S = \frac{1}{4}\log N + \mathcal{O}(1)$ for large $N$. In the present model, by the Perron--Frobenius theorem, the largest eigenvalue is given by $\Lambda=4(N-1)\eta(1-\eta)m^2=\mathcal{O}(N)$. Therefore, Eq. (\ref{eq ts correspond to continuous}) also reproduces the result of \cite{achilli2026theory} at the level of the leading scaling with $N$. This indicates that, for the special case of Ising variables with two classes distinguished by opposite first moments $\pm m$, the order of the speciation time can be estimated from a different route, namely as a second-order transition obtained by a high-temperature expansion.

Also, since $t > 0$, Eq. (\ref{eq ts}) implies that speciation occurs when $\Lambda > 1$. By applying the Perron–Frobenius theorem together with Eq. (\ref{def P0}), this condition can be written as
\begin{eqnarray}
4(N-1)\eta(1-\eta)m^{2} > 1.
\end{eqnarray}
This expression implies that trivial cases in which speciation does not occur are given by $\eta = 0, 1$ and $m = 0$.

Similarly, the maximum eigenvalue of the covariance matrix of $P_{0}(\bm x_{0})$ is given by
\begin{eqnarray}
    \Lambda_{\mathrm{cov}} = \left( 1-(2\eta - 1)^{2}m^{2}\right) + 4(N-1)\eta(1-\eta)m^{2}\label{eq max eigenvalue of P0}.
\end{eqnarray}
The first term of Eq. (\ref{eq max eigenvalue of P0}) can be neglected because the second term is of order $\mathcal{O}(N)$. Hence, for a simplicity, we use $\Lambda_{\mathrm{cov}}$ as $\Lambda$ for the numerical and the real-data experiments shown later.

\subsection{The collapse time}
\label{subsec:sec4_2}

Collapse is the moment when the trajectory of the generated data finds the data sample closest to itself.
This collapse situation can be formulated, by following \cite{biroli2024dynamical}, as a relation between two Shannon entropy densities. The first is the entropy density of the marginal distribution $P_{t}(\bm x_{t})$, defined as $S(t) = -\frac{1}{N}\sum_{\bm x_{t}} P_{t}(\bm x_{t}) \log P_{t}(\bm x_{t})$. The second is the entropy density of a distribution concentrated on individual training data point in a well-separated manner, given by $S^{sep}(t) = \frac{\log p}{N} - \frac{1 + \theta^{t}}{2}\log \frac{1 + \theta^{t}}{2} - \frac{1 - \theta^{t}}{2}\log \frac{1 - \theta^{t}}{2}$, where $p$ denotes the number of data.

The former, namely the Shannon entropy density of $P_{t}(\bm x_{t})$, provides the correct entropy density at least until collapse occurs. Once collapse takes place, however, since collapse corresponds to the trajectory capturing a training data point, the latter, the Shannon entropy density of the well-separated distribution over the data points, can well approximate the exact entropy. In other words, the collapse time $t_{C}$ is precisely the moment at which the entropy of the marginal distribution $P_{t}(\bm x_{t})$ transitions to that of the well-separated distribution. Thus, the criterion that the collapse time $t_{C}$ should satisfy is given by
\begin{eqnarray}
    S(t_{C}) = S^{sep}(t_{C}).\label{def general collapse criterion}
\end{eqnarray}

Throughout this collapse analysis, we write the formulas for the constant noise schedule. The extension to a time-dependent noise schedule is obtained by replacing $\theta^t$ with the cumulative signal $\Theta_t=\prod_{s=1}^{t}(1-\beta_s)$.

To calculate $S(t)$, we rewrite $P_{t}(\bm x_{t})$ in terms of $m_{t} = \frac{1}{N}\sum_{i=1}^{N} x_{ti}$ and the number of components satisfying $x_{ti} = 1$, denoted by $s = \frac{N(1+m_{t})}{2}$. For a given configuration $\bm x_{t}$, the value of $P_{t}(\bm x_{t})$ is uniquely determined by $m_{t}$ (or equivalently $s$). We then define the corresponding probability as $p_{t}(s) = 2^{-N}\left[ \eta(1 + m\theta^{t})^{s}(1 - m\theta^{t})^{N-s} + (1-\eta)(1 - m\theta^{t})^{s}(1 + m\theta^{t})^{N-s} \right]$. Since the number of configurations with the same probability is ${N \choose s}$, the entropy can be written as $S(t) = -(1/N)\sum_{s=0}^{N} {N \choose s}p_{t}(s)\log p_{t}(s)$.

However, for real data, $P_{0}$ is unknown, and consequently the marginal distribution $P_{t}$ is also unknown. We therefore derive an explicit expression for the empirical marginal distribution $P_{t}^{e}$ that depends on the observed data. We denote $\mathcal D = \{ \bm x^{\mu} \}_{\mu = 1}^{p}$ the entire dataset, where each data point is $\bm x^{\mu} \in \{-1, 1\}^{N}$ for all data indices $\mu = 1,2,\cdots,p$. In this case, the data distribution becomes the following conditional distribution:
\begin{eqnarray}
    P_{0}^{e}(\bm x_{0} | \mathcal{D}) = \frac{1}{p}\sum_{\mu = 1}^{p}\prod_{i = 1}^{N} \delta(x_{0i} - x_{i}^{\mu}).
\end{eqnarray}
Then, the marginal distribution at each time that depends on the dataset $\mathcal D$, $P_{t}^{e}(\bm x_{t} | \mathcal{D})$ is obtained as follows:
\begin{eqnarray}
    P_{t}^{e}(\bm x_{t} | \mathcal{D}) &=& \sum_{\bm x_{0}}P(\bm x_{t} | \bm x_{0})P_{0}^{e}(\bm x_{0} | \mathcal{D})\\
                        &=& \sum_{\bm x_{0}} \prod_{i = 1}^{N}\frac{1 + \theta^{t} x_{ti}x_{0i}}{2} \cdot \frac{1}{p}\sum_{\mu = 1}^{p}\prod_{i = 1}^{N}\delta(x_{0i} - x_{i}^{\mu})\\
                        &=& \frac{1}{p}\sum_{\mu = 1}^{p}\prod_{i=1}^{N}\frac{1 + \theta^{t} x_{ti}x_{i}^{\mu}}{2}\label{eq Pt_e non int ising}\\
                        &=&
                        \frac{1}{p} \sum_{\mu}^{p} \prod_{i = 1}^{N} \frac{e^{F_{t}  x_{ti} x_{i}^{\mu}}}{2\cosh F_{t}}\label{eq Pt_e exp form}\\
                        &=&
                        \frac{1}{p} \frac{1}{[2\cosh F_{t}]^{N}} \left( \sum_{\mu = 1}^{p} e^{F_{t} \bm x_{t}\cdot \bm x^{\mu}}\right).                  
\end{eqnarray}
From Eq. (\ref{eq Pt_e non int ising}) to Eq. (\ref{eq Pt_e exp form}), we have used Eq. (\ref{eq def binom to exp}). Then, the collapse criterion Eq. (\ref{def general collapse criterion}) becomes as $S^{e}(t) = S^{sep}(t)$, where $S^{e}(t) = -\frac{1}{N} P_{
t}^{e}(\bm x_{t} | \mathcal{D}) \log P_{t}(\bm x_{t}|\mathcal{D})$. However, the direct evaluation of $S^{e}(t)$ is also difficult. We thus compute the following empirical average:
\begin{eqnarray}
    S^{e}(t) \approx -\frac{1}{n_{sample}N} \sum_{\nu = 1}^{n_{sample}} \log P_{t}^e(\boldsymbol x_{t}^{(\nu)} | \mathcal{D})\label{eq def S_e_t},
\end{eqnarray}
where $n_{sample}$ is the sample size.

The same conclusion obtained from this information-theoretic detection of collapse can also be derived by interpreting the phenomenon as a phase transition in the statistical mechanics of disordered systems, the condensation transition of the Random Energy Model \cite{derrida1981random} (REM). The REM-based analysis of collapse in Diffusion Models was first carried out by \cite{biroli2024dynamical}, while closely related analyses had previously been performed in the context of dense associative memory models \cite{lucibello2024exponential}. This correspondence is theoretically significant in that the collapse can be identified with the REM condensation transition. Moreover, it is also of practical importance: in realistic settings with high dimensionality and large data sizes, the accurate computation of $S^{e}(t)$ becomes computationally prohibitive, whereas the collapse time $t_{C}$ can be efficiently estimated via the REM-based analysis. The derivation of the collapse time based on the REM proceeds as follows.

We divide the the factor $Z = \sum_{\mu = 1}^{p} e^{F_{t}\bm x_{t} \cdot \bm x^{\mu}}$ into two partition functions:
\begin{eqnarray}
    Z &=& Z_{+} + Z_{-},
\end{eqnarray}
where,
\begin{eqnarray}
    Z_{+} &=& \sum_{\mu \in +} e^{F_{t} \bm x_{t} \cdot \bm x^{\mu}},
    \qquad 
    Z_{-} = \sum_{\mu \in -} e^{F_{t} \bm x_{t} \cdot \bm x^{\mu}}.
\end{eqnarray}
The two symbols $+$ and $-$ represent the set of classes $+$ and the set of classes $-$ of all data indices. In the following, we denote by $p_{+}$ and $p_{-}$ the numbers of data points belonging to the $+$ and $-$ classes, respectively. The data indices are then assigned such that $\mu = 1, \cdots ,p_{+}$ correspond to data points in the $+$ class, while $\mu = p_{+} + 1, \cdots , p$ correspond to those in the $-$ class. 

From here, we focus on $Z_{+}$. We assume the first data $\bm x^{1}$ denotes the most closest data at the collapse. Then, the partition function $Z_{+}$ can be divided into two parts: 
\begin{eqnarray}
    Z_{+} = Z_{1} + Z_{2...p_{+}},
\end{eqnarray}
where,
\begin{eqnarray}
    Z_{2...p_{+}} = \sum_{\mu = 2}^{p_{+}} e^{F_{t} \bm x_{t} \cdot \bm x^{\mu}}.
\end{eqnarray}
When the closest data point belongs to the negative class, that is, when the same analysis is carried out with respect to $Z_{-}$, the final result is reproduced exactly.
The reason is straightforward: compared to the analysis of $Z_{+}$ presented below, considering $Z_{-}$ only induces sign reversals in several parameters. However, the criterion that determines the collapse time through the REM remains invariant under these sign reversals.
This point will be explicitly verified again in the following analysis. 

Because the collapse is the moment when generated data in backward process achieves $\bm x^{1}$, its criterion in the REM approach is then
\begin{eqnarray}
    Z_{1} = Z_{2...p_{+}}\label{eq def REM collapse criterion}.
\end{eqnarray}
The inner product $\bm x_{t} \cdot \bm x^{{1}}$ can be approximated by $\bm x_{t} \cdot \bm x^{{1}} \approx N\theta^{t}$. Then the first part of the partition function is given by $Z_{1} \approx \exp(F_{t}N\theta^{t})$. The latter part, $Z_{2...p_{+}}$ can be calculated through the REM. 

As a result, we obtain 
\begin{eqnarray}
    Z_{2...p_{+}} = e^{N[s_{t} + F_{t}\theta^{t}]},\label{eq def partition func Z2_p main text}
\end{eqnarray}
where $s_{t}$ is, using the parameter $\alpha:= \frac{\log p}{N}$ and the representation of the Kullback-Leibler divergence between two binary distributions: $D_{\mathrm{KL}}(x||y) = x\log\frac{x}{y} + (1-x)\log\frac{1-x}{1-y}$, given as follows
\begin{eqnarray}
    \hspace{-10mm}s_{t} =
\alpha - \frac{1 + m}{2}
D_{\mathrm{KL}}\!\left(
\frac{1 + \theta^{t}}{2}
\,\middle\|\,
\frac{1 + m\theta^{t}}{2}
\right)
- \frac{1 - m}{2}
D_{\mathrm{KL}}\!\left(
\frac{1 + \theta^{t}}{2}
\,\middle\|\,
\frac{1 - m\theta^{t}}{2}
\right)
\label{eq st main text}.
\end{eqnarray}
Here, $s_{t}$ is the microcanonical entropy density of the system when the microscopic energy is
\begin{eqnarray}
    \varepsilon = \frac{1}{N} \sum_{i = 1}^{N}x_{ti}x_{i}^{\mu}.
\end{eqnarray}
The derivation of Eq. (\ref{eq def partition func Z2_p main text}) and Eq. (\ref{eq st main text}) is done by considering the conditional distribution $P(\varepsilon|\bm x_{t})$ of the microscopic energy $\varepsilon$ given the sample $\bm x_{t}$ (for details, see Appendix \ref{app:appB_rev}).

In the REM, the point at which the microcanonical entropy vanishes corresponds to a condensation transition. At this point, the contribution to the partition function is dominated almost entirely by the ground state.
In the context of the present backward diffusion dynamics, this situation can be understood as the time at which the partition function $Z_{2..p_{+}} = Z_{2...p_{+}}(\bm x_{t})$ at each step of the reverse diffusion process can be effectively dominated by the closest data point, that is, the time at which collapse occurs. Using the criterion of the collapse in the REM Eq. (\ref{eq def REM collapse criterion}), $Z_{1} \approx e^{NF_{t}\theta^{t}}$, and Eq. (\ref{eq st main text}), it is clear that the condensation transition condition $s_{t} = 0$ is equivalent to the criterion in Eq. (\ref{eq def REM collapse criterion}). Therefore, within the REM framework, the collapse time is determined as the numerical solution with respect to $t$ of the equation $s_{t} = 0$, where $s_{t}$ is given by Eq. (\ref{eq st main text}).

Having established the collapse condition, we can also derive the criterion for avoiding the curse of dimensionality. The curse of dimensionality refers to the growth of the number of samples needed to cover the typical set of the data distribution as the raw data dimension $N$ increases. Since the curse of dimensionality is a fundamental issue in understanding the performance of diffusion models, we now provide a simple analysis of this problem in the context of discrete generative diffusion. From Eq. (\ref{eq st main text}), at $t=0$ we obtain
\begin{eqnarray}
s_{t=0} &=& \alpha + \frac{1+m}{2}\log\frac{1+m}{2} + \frac{1-m}{2}\log\frac{1-m}{2}\\
&=&\alpha - H_{\mathrm B}\!\left(\frac{1+m}{2}\right),
\end{eqnarray}
where $H_{\mathrm B}(x)=-x\log x-(1-x)\log(1-x)$ denotes the binary entropy. Since the collapse time is determined by the condition $s_t=0$, avoiding collapse before the end of the reverse process requires $\alpha > H_{\mathrm B}((1+m)/2)$, or equivalently
\begin{eqnarray}
p > \exp\left[N H_{\mathrm B}\!\left(\frac{1+m}{2}\right)\right].
\end{eqnarray}
Thus, for any $0 \leq m<1$, the required number of data samples grows exponentially with the raw data dimension $N$. This is analogous to the continuous case \cite{biroli2024dynamical} and shows that the curse of dimensionality also appears for discrete data. The case $m=1$ corresponds to the most extreme situation in which the positive and negative classes consist only of $(1,\ldots,1)$ and $(-1,\ldots,-1)$, respectively. Since $H_{\mathrm B}((1+m)/2)=0$, the condition reduces to $p>1$. In this limit, each class has only one configuration, so concentration onto that configuration is indistinguishable from correctly reproducing the data distribution.

\section{Numerical validation}\label{sec:sec5}

\subsection{Numerical validation of the speciation time}\label{subsec:sec5_1}

We here show the results of comparison between the theoretical prediction of $t_{S}$ by Eq. (\ref{eq ts}) and the bifurcation of the trajectories of the reverse process of the effective model that was introduced in Section 2.

Fig. 1 shows class-balanced and class-imbalanced cases with $N = 10000$. The vertical axis label is defined as $m_{t} = (1/N)\sum_{i = 1}^{N} x_{ti}$. The parameter $m$ is set to 0.8 in both cases, and the plotted trajectories therefore approach $m_{0} = \pm 0.8$. The theoretical prediction of $t_{S}$, calculated from Eq. (\ref{eq ts}) and indicated by the black vertical dashed line, accurately captures the bifurcation in the bundles of trajectories for each case.

\begin{figure*}[tb]
\centering

\begin{minipage}{0.49\linewidth}
    \centering
    \includegraphics[width=\linewidth]{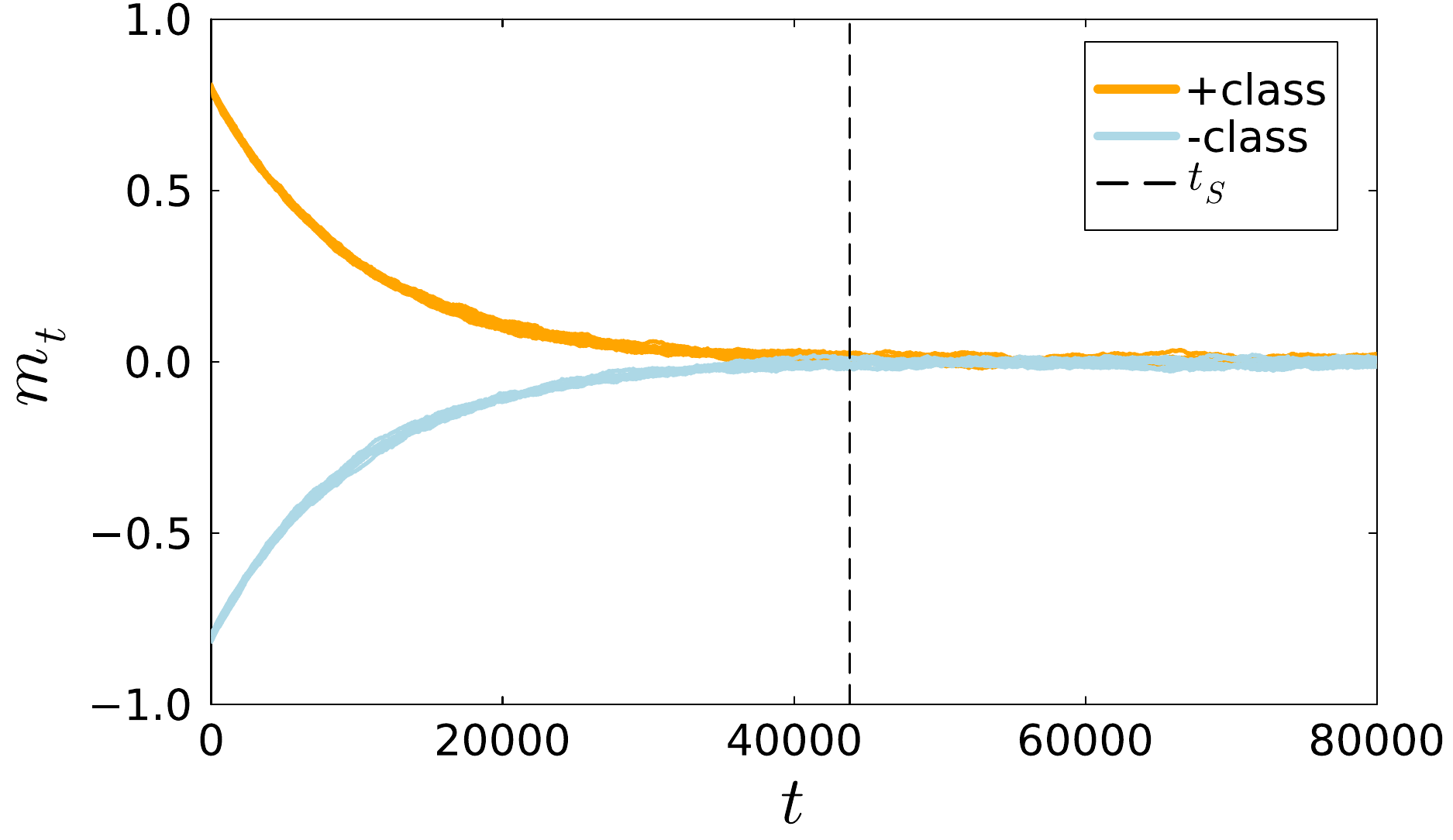}\\
    (a) $\eta = 0.5$
\end{minipage}
\hfill
\begin{minipage}{0.49\linewidth}
    \centering
    \includegraphics[width=\linewidth]{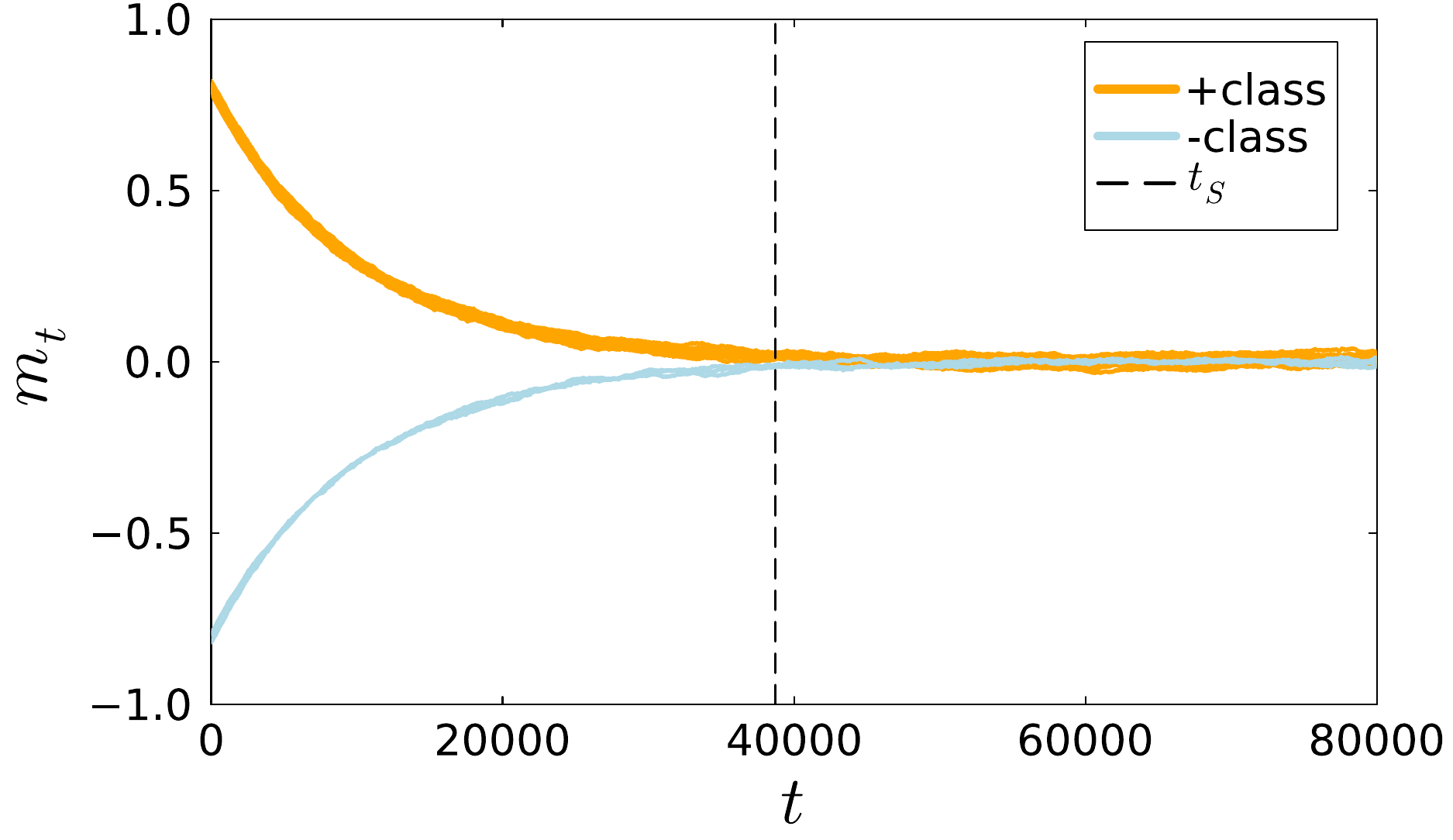}\\
    (b) $\eta = 0.9$
\end{minipage}

\caption{Comparison of the results for class-balanced setting and class-imbalanced setting of $t_{S}$ with $N = 10000$. We have set $m = 0.8$ and $\beta = 10^{-4}$ in both cases. The orange plots represent trajectories with $m_{t} > 0$ at $t = 0$, while the light-blue plots represent those with $m_{t} < 0$ at $t = 0$. The number of displayed trajectories is 20 in total, combining the positive and negative classes for both the class-balanced and class-imbalanced settings. The black vertical dashed line indicates the value of $t_{S}$ computed from Eq. (\ref{eq ts}). (a): The reverse process dynamics of class-balanced setting with $\eta = 0.5$. (b): Class-imbalanced setting with $\eta = 0.9$. }
\label{fig:comparison}
\end{figure*}

To more precisely test the validity of the theoretical prediction of $t_{S}$, we use the cloning method. What we aim to evaluate here is the probability that two data generation trajectories, which share exactly the same configuration at a given time $t$ in the reverse diffusion process, belong to the same class at time $t = 0$. 

From its definition, the cloning probability introduced here can be regarded as an order parameter that characterizes how the trajectories of the generated data capture the global class structure in the reverse diffusion process. By the definition of the speciation time, $t_{S}$ is expected to coincide with the transition point at which the cloning probability exhibits phase-transition-like behavior. In the following analysis, we numerically verify this correspondence. The following explanation of the cloning method is also based on the description of the cloning method for continuous data given in the previous work by \cite{biroli2024dynamical}.

In the reverse process, we consider two trajectories, $\bm x_{t}^{(1)}$ and $\bm x_{t}^{(2)}$, that share exactly the same configuration $\bm y$ at a given time $t$. By construction, we have $\bm x_{t}^{(1)} = \bm x_{t}^{(2)} = \bm y$. We denote by $p(\bm x^{(1)}, 0 | \bm y, t)$ and $p(\bm x^{(2)}, 0 | \bm y, t)$ the probabilities that the generated data, being $\bm y$ at time $t$, evolve into $\bm x^{(1)}$ and $\bm x^{(2)}$ at time $0$, respectively. We further denote by $p(\bm y, t)$ the probability that the generated data takes the value $\bm y$ at time $t$. Under these definitions, the probability that the two generated data $\bm x^{(1)}$ and $\bm x^{(2)}$, which coincide at time $t$ as $\bm y$ belong to the same class at time $0$, $q(\bm y, t)$ is given by

\begin{eqnarray}
    q(\bm y, t) = \sum_{\bm x^{(1)}, \bm x^{(2)} \atop m_{0}^{(1)}\times m_{0}^{(2)} > 0} p(\bm x^{(1)}, 0 | \bm y, t)p(\bm x^{(2)}, 0 | \bm y, t) p(\bm y, t),
\end{eqnarray}
where $m_{0}^{(1)} = (1/N) \sum_{i = 1}^{N}x_{0i}^{(1)}$, $m_{0}^{(2)} = (1/N) \sum_{i = 1}^{N}x_{0i}^{(2)}$. In order to evaluate $\phi(t)$, we calculate $p(\bm x^{(1)}, 0 | \bm y, t)$ and $p(\bm x^{(2)}, 0 | \bm y, t)$. By the formula for conditional probability, $p(\bm x^{(1)}, 0 | \bm y, t)$ can be formally rewritten as follows.
\begin{eqnarray}
    p(\bm x^{(1)}, 0 | \bm y, t) &=& p(\bm x^{(1)}, 0 ; \bm y, t) /p(\bm y, t)\\
    &=& \frac{p(\bm y, t | \bm x^{(1)}, 0) p(\bm x^{(1)},0)}{p(\bm y, t)}\label{eq clone prob 1 line 2}\\
    &=& \frac{P(\bm y | \bm x_{0}^{(1)}) P_{0}(\bm x^{(1)})}{P_{t}(\bm y)}\label{eq clone prob 1 line 3}.
\end{eqnarray}
Where $\bm x_{0}^{(1)}$ denotes the generated data of the trajectory $\bm x^{(1)}$ at $t = 0$. From Eq. (\ref{eq clone prob 1 line 2}) to Eq. (\ref{eq clone prob 1 line 3}), we use the forward process distribution Eq. (\ref{eq forward 0tot}) for $p(\bm y,t | \bm x^{(1)}, 0)$, the data distribution Eq. (\ref{def P0}) for $p(\bm x^{(1)}, 0)$, and the marginal distribution Eq. (\ref{eq def Pt}) for $p(\bm y, t)$. Based on the same procedure, we get
\begin{eqnarray}
    p(\bm x^{(2)}, 0 | \bm y, t) = \frac{P(\bm y | \bm x_{0}^{(2)}) P_{0}(\bm x^{(2)})}{P_{t}(\bm y)}.
\end{eqnarray}

Therefore, one can write the probability $q(\bm y, t)$ as follows:
\begin{eqnarray}
 q(\bm y, t) = \sum_{\bm x^{(1)}, \bm x^{(2)} \atop m_{0}^{(1)}\times m_{0}^{(2)} > 0} \frac{1}{P_{t}(\bm y)} \left( \prod_{i = 1}^{N} \frac{1 + \theta^{t}y_{i}x_{0i}^{(1)}}{2} \right) \left( \prod_{i = 1}^{N} \frac{1 + \theta^{t}y_{i}x_{0i}^{(2)}}{2} \right) \nonumber\\
    \times \left( \eta \prod_{i = 1}^{N} \frac{1 + mx_{0i}^{(1)}}{2} +  (1-\eta) \prod_{i = 1}^{N} \frac{1 - mx_{0i}^{(1)}}{2} \right)\nonumber\\
    \times
    \left( \eta \prod_{i = 1}^{N} \frac{1 + mx_{0i}^{(2)}}{2} +  (1-\eta) \prod_{i = 1}^{N} \frac{1 - mx_{0i}^{(2)}}{2} \right)\label{eq def cloning prob kind}\\
    =
    \sum_{\bm x^{(1)}, \bm x^{(2)}} \frac{1}{P_{t}(\bm y)} \left( \prod_{i = 1}^{N} \frac{1 + \theta^{t}y_{i}x_{0i}^{(1)}}{2} \right) \left( \prod_{i = 1}^{N} \frac{1 + \theta^{t}y_{i}x_{0i}^{(2)}}{2} \right) \nonumber\\
    \times
    \left( \eta^{2} \prod_{i = 1}^{N} \frac{1 + mx_{0i}^{(1)}}{2} \frac{1 + mx_{0i}^{(2)}}{2} + (1-\eta)^{2} \prod_{i = 1}^{N} \frac{1 - mx_{0i}^{(1)}}{2} \frac{1 - mx_{0i}^{(2)}}{2} \right)\label{eq def cloning prob kind no cross term}\\
    = \frac{1}{P_{t}(\bm y)} \left[ \left( \eta \prod_{i = 1}^{N}\frac{1 + m\theta^{t} y_{i}}{2}\right)^{2} + \left( (1-\eta) \prod_{i = 1}^{N}\frac{1 - m\theta^{t} y_{i}}{2}\right)^{2}\right]\label{eq result of cloning prob kind}.
\end{eqnarray}
In the transformation from Eq. (\ref{eq def cloning prob kind}) to Eq. (\ref{eq def cloning prob kind no cross term}), we used the condition that the two trajectories $\bm x^{(1)}$ and $\bm x^{(2)}$ belong to the same class at time $0$ by excluding the two cross terms between the $+$ class and the $-$ class in the product $P_{0}(\bm x^{(1)}) P_{0}(\bm x^{(2)})$. Furthermore, in the step from Eq. (\ref{eq def cloning prob kind no cross term}) to Eq. (\ref{eq result of cloning prob kind}), the summations over $\bm x^{(1)}$ and $\bm x^{(2)}$ are carried out independently for each index $i$.

The cloning probability $\phi(t)$ is obtained by marginalizing $q(\bm y,t)$ over $\bm y$ as follows.
\begin{eqnarray}
    \phi(t) &=& \sum_{\bm y} q(\bm y, t)\label{eq def cloning prob}\\
    &=&
    \sum_{\bm y} \frac{1}{P_{t}(\bm y)} \left[ \left( \eta \prod_{i = 1}^{N}\frac{1 + m\theta^{t} y_{i}}{2}\right)^{2} + \left( (1-\eta) \prod_{i = 1}^{N}\frac{1 - m\theta^{t} y_{i}}{2}\right)^{2}\right]\label{eq cloning prob line 2}\\
    &=&
    \sum_{\bm y} \frac{ \left( \eta \prod_{i = 1}^{N}\frac{1 + m\theta^{t} y_{i}}{2}\right)^{2} + \left( (1-\eta) \prod_{i = 1}^{N}\frac{1 - m\theta^{t} y_{i}}{2}\right)^{2}
    }{\eta\prod_{i = 1}^{N}\frac{1 + \theta^{t}m y_{i}}{2} + (1-\eta) \prod_{i = 1}^{N}\frac{1 - \theta^{t}m y_{i}}{2}
    }\label{eq cloning prob line 3}.
\end{eqnarray}
However, in its present form, carrying out the summation in Eq. (\ref{eq clone prob 1 line 3}) is computationally intractable. Nevertheless, inspection of the numerator and denominator in Eq. (\ref{eq clone prob 1 line 3}) reveals that each term depends only on the ``magnetization" of $\bm y$, $\sum_{i=1}^{N} y_{i}$, and not on the specific configuration. Therefore, $\phi(t)$ can be expressed in terms of a binomial distribution over the number of components satisfying $y_{i} = 1$, namely
$k_{+} = (N + \sum_{i=1}^{N} y_{i})/2$.
Accordingly, by replacing the summation $\sum_{\bm y}$ with $\sum_{k_{+}=0}^{N}$, we obtain
\begin{eqnarray}
    \hspace{-10mm}\phi(t) = \sum_{k_{+} = 0}^{N} \frac{\left( \eta \hspace{1mm} Bin\!\left(
k_{+}| N, \frac{1 + m\theta^{t}}{2} \right)
\right)^{2} + \left( (1-\eta) \hspace{1mm} Bin\!\left(
k_{+} |N, \frac{1 + m\theta^{t}}{2} \right)
\right)^{2}
    }
    { \eta \hspace{1mm} Bin\!\left(
k_{+} | N, \frac{1 + m\theta^{t}}{2} \right)
 + (1-\eta) \hspace{1mm} Bin\!\left(
k_{+} | N, \frac{1 + m\theta^{t}}{2} \right)
    }\label{eq cloning prob binary dist},
\end{eqnarray}
where $Bin(k | N, \theta)$ is a Binomial probability distribution function for the integer $k \geq 0$, with number of trials $N$ and success probability $\theta$. 

\begin{figure*}[t]
\centering

\begin{minipage}{0.49\linewidth}
    \centering
    \includegraphics[width=\linewidth]{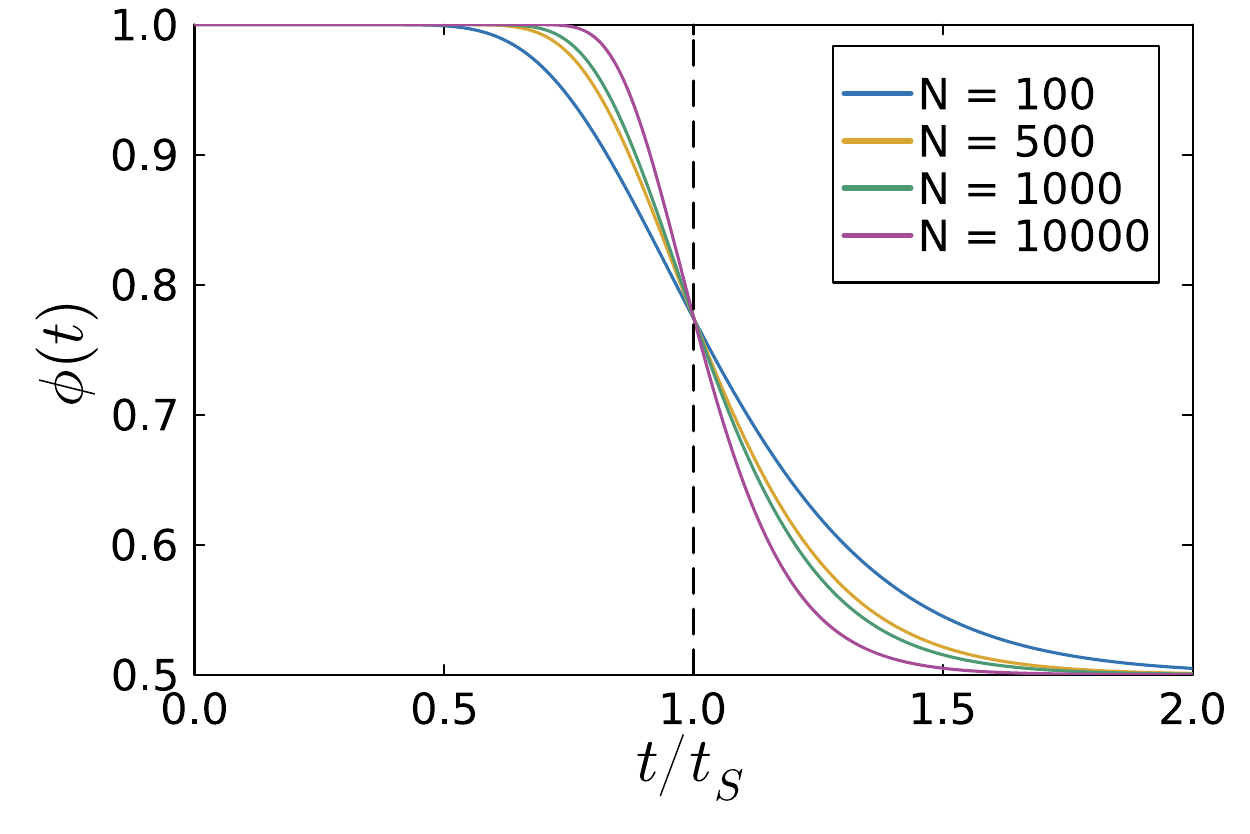}\\
    (a) $\eta = 0.5$
\end{minipage}
\hfill
\begin{minipage}{0.49\linewidth}
    \centering
    \includegraphics[width=\linewidth]{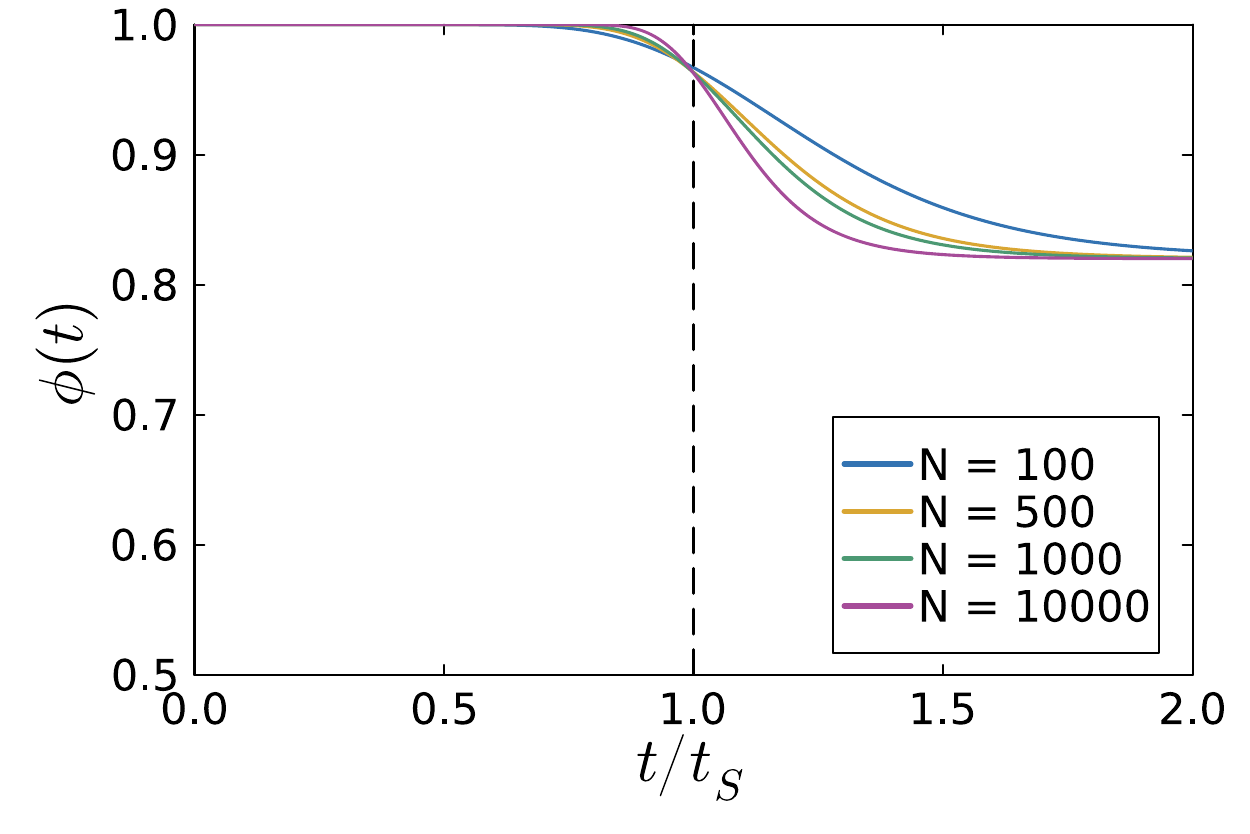}\\
    (b) $\eta = 0.9$
\end{minipage}

\caption{Cloning probability $\phi(t)$ plotted as a function of the rescaled time $t/t_{S}$ for increasing system sizes. For both cases we have set $m = 1$ and $\beta = 10^{-4}$.
(a) Results for the balanced data case with $\eta = 0.5$. The intersection value of the four $\phi(t)$ curves is 0.770.
(b) Results for the imbalanced data case with $\eta = 0.9$. The intersection value of the four $\phi(t)$ curves is 0.967.}
\label{fig:comparison}
\end{figure*}
The direct numerical evaluation of Eq. (\ref{eq cloning prob binary dist}) requires computing two binomial probabilities $N$ times at each time step $t$, and is therefore significantly faster than the direct evaluation of Eq. (\ref{eq cloning prob line 3}).

Fig. 2 shows the results of the numerical evaluation of $\phi(t)$ for $\eta = 0.5$ and $\eta = 0.9$ for increasingly larger dimensions. The horizontal axis is the reverse time step rescaled by $t_{S}$. For both cases we have set $m = 1$ and $\beta = 10^{-4}$. The speciation time $t_{S}$ is already obtained as Eq. (\ref{eq ts}). 

As shown in Fig. 2, for all dimensions $N$, the cloning probability $\phi(t)$ almost identical at $t = t_{S}$ for both $\eta = 0.5$ and $\eta = 0.9$ cases. The figure suggests a step-function-like increase of $\phi(t)$ at $t = t_{S}$ in the infinite-dimensional limit. Therefore, based on finite-size scaling theory \cite{privman1990finite}, $t_{S}$ can be regarded as the transition point of the cloning probability $\phi(t)$.

\subsection{Numerical validation of the collapse time}\label{subsec:sec5_2}

Here we show a numerical validation of the theoretical prediction of the collapse time, which is the solution of $s_{t} = 0$ where $s_{t}$ is given as Eq. (\ref{eq st main text}). We denote the solution of $s_{t} = 0$ with respect to $t$, $t_{C}^{REM}$. We first check the validity of $t_{C}^{REM}$ by comparing the time that satisfies $S(t) = S^{sep}(t)$. To do this, we define the entropy difference normalized by $\alpha = \frac{\log p}{N}$, $\Delta S(t):= (S^{sep}(t) - S(t))/\alpha$. When $t \gg 1$, $\theta^{t} \approx 0$ hence the configurational entropy of $S^{sep}(t)$, $-\frac{1 + \theta^{t}}{2}\log\frac{1 + \theta^{t}}{2} - \frac{1 - \theta^{t}}{2}\log \frac{1 - \theta^{t}}{2}$ and $S(t)$ both becomes the entropy density of the Bernoulli distribution with equal probability $\log 2$. Thus, $\Delta S(t) = 1$ when $t \gg 1$. 
\begin{figure*}[t]
\centering
\begin{minipage}{0.49\linewidth}
    \centering
    \includegraphics[width=\linewidth]{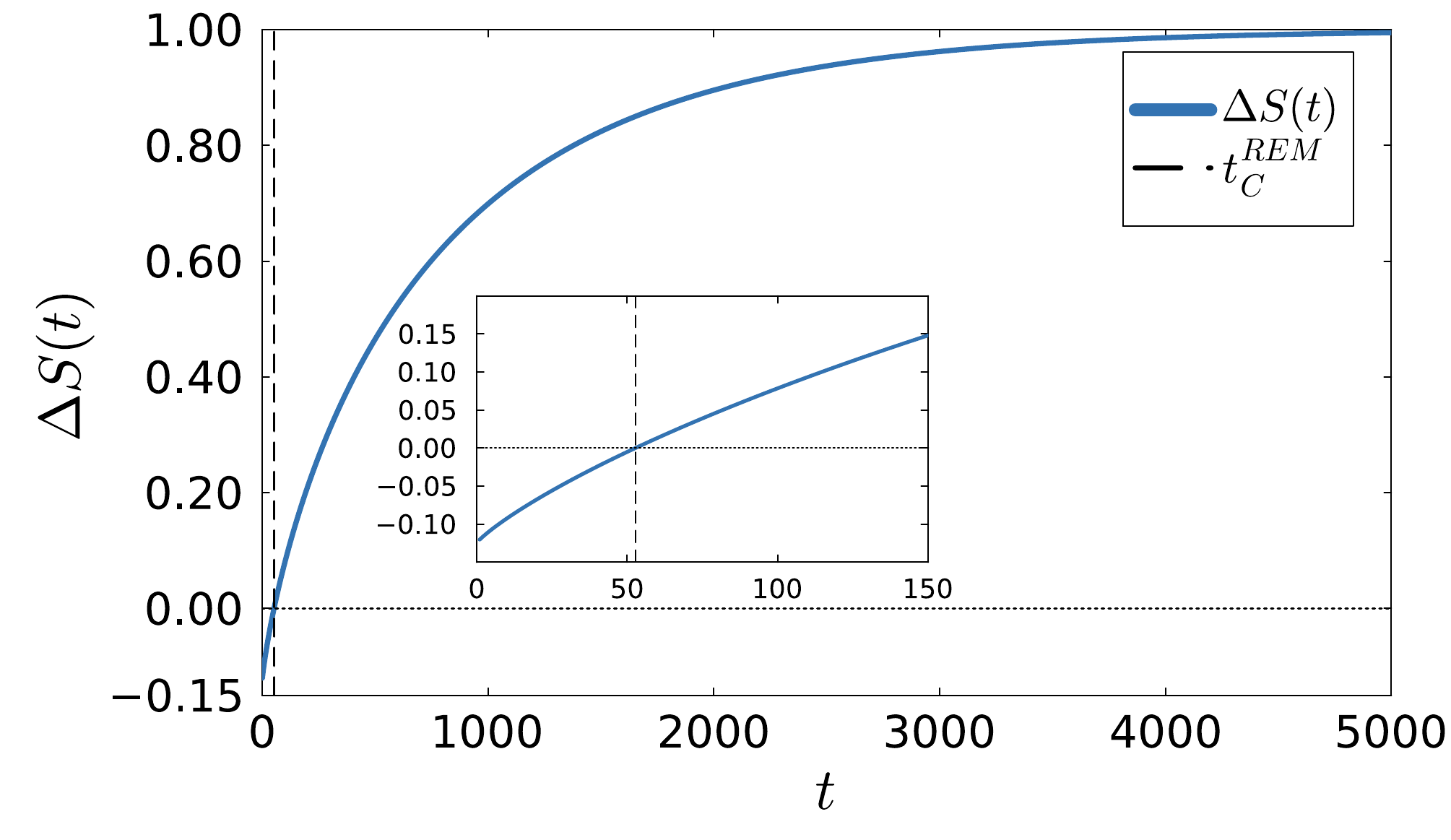}\\
    (a) $t$ v.s. $\Delta S(t)$ with $t_{C}^{REM}$ ($\alpha = 0.5$)
\end{minipage}
\hfill
\begin{minipage}{0.49\linewidth}
    \centering
    \includegraphics[width=\linewidth]{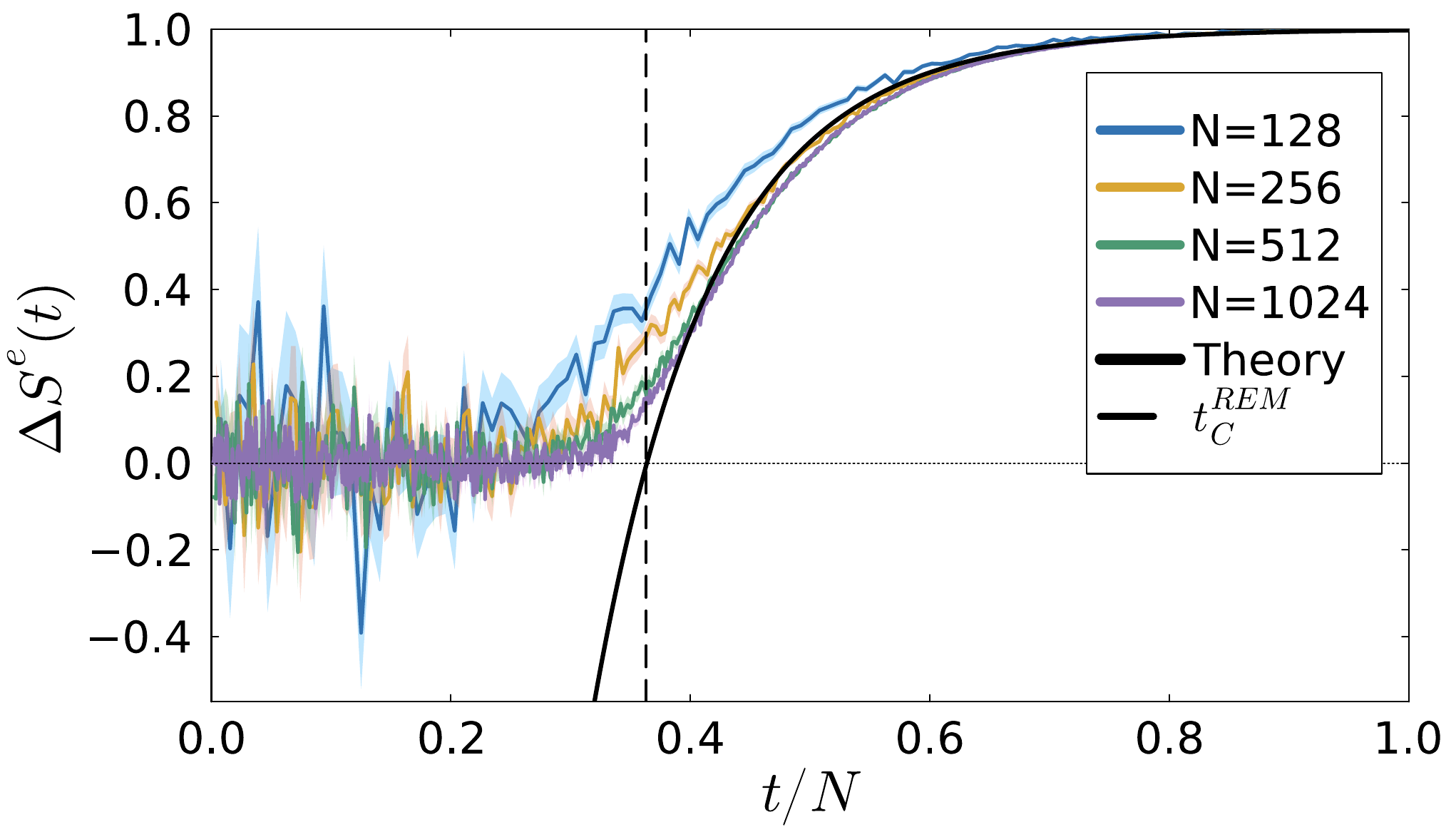}\\
    (b) $t/N$ v.s. $\Delta S^{e}(t)$ with $t_{C}^{REM}$ ($\alpha = 0.01$)
\end{minipage}
\caption{Comparison between the collapse time $t_{C}^{REM}$, obtained as the numerical solution of $s_{t} = 0$, and the general criterion Eq. (\ref{def general collapse criterion}), for both the theoretical and empirical values of the Shannon entropy density $S(t)$. For all cases we set $m = 0.5$, $\eta = 0.5$, and $\beta = 5\times10^{-4}$(a): Comparison between $\Delta S(t)$ and the collapse time $t_{C}^{REM}$ obtained as the numerical solution of $s_{t} = 0$.
The solid navy-blue curve shows $\Delta S(t)$, while the dashed vertical line indicates $t_{C}^{REM}$.
The parameters are set to $N = 10000$ and $\alpha = 0.5$. (b): Finite-size scaling of the empirical entropy difference $\Delta S^{e}(t)$ at fixed $\alpha = 0.01$.
The horizontal axis is the rescaled time $t/N$, chosen so that all curves overlap.
Colored curves represent $\Delta S^{e}(t)$ for increasing system sizes $N$, while the curve labeled “Theory” shows the theoretical prediction $\Delta S(t)$ evaluated at $N = 10000$.
The dashed vertical line denotes the collapse time $t_{C}^{REM}$, given by the solution of $s_{t} = 0$ for the same parameters. Each ribbon indicates the standard error.}
\label{fig:comparison}
\end{figure*}
Fig. 3 (a) shows the result. For this figure, we set $N = 10000$ and $\alpha = 0.5$ for the calculation of $\Delta S(t)$ and $t_{C}^{REM}$. It can be seen that $t_{C}^{REM}$, represented by the dashed vertical line, captures well the timing at which the backward dynamics of $\Delta S(t)$, represented by the navy blue line, crosses zero. The REM analysis also accurately detects the collapse time even for discrete data.

Second, with real-data experiments in mind, we examine whether $S^{e}(t)$ provides a valid approximation to $S(t)$.
This analysis serves as a test of whether the REM analysis can correctly detect the collapse time in real-data experiments, once finite-size effects are neglected.
We investigate whether the empirical entropy difference $\Delta S^{e}(t) := (S^{sep}(t) - S^{e}(t))/\alpha$ converges to $\Delta S(t)$ as the dimension $N$ is increased while keeping $\alpha = \frac{\log p}{N}$ fixed. Fig. 3 (b) shows the result. We set $\alpha = 0.01$ due to the efficiency for the computation of $\Delta S^{e}(t)$. The horizontal axis is the rescaled time $t/N$, chosen so that all curves overlap. The colored lines represent the curves of $\Delta S^{e}(t)$ for increasing dimensions $N$.
The curve labeled ``Theory" corresponds to $\Delta S(t)$ evaluated at $N = 10000$ and $\alpha = 0.01$.
The quantity $t_{C}^{REM}$, also represented by the dashed vertical line, is the solution of $s_{t} = 0$ for the same set of parameters.

From this figure, it is reasonable to expect that, for $t \ge t_{C}$, the discrepancy between $\Delta S^{e}(t)$ and its theoretical value $\Delta S(t)$ is due only to finite-size effects.
We next comment on the difference between $\Delta S^{e}(t)$ and $\Delta S(t)$ in the regime $t \leq t_{C}$.
In this regime, the data-generation distribution becomes sharply concentrated around each data point, effectively resembling a set of delta function-like peaks, and thus $S^{e}(t)$ coincides with $S^{sep}(t)$.
By contrast, $S(t)$ is the entropy of $P_{t}(\bm x_{t})$, which is defined in the limit of an infinite number of data points $p \rightarrow \infty$ and hence does not reflect the individuality of each finite training data.
Therefore, once the dependence of empirical data becomes very strong for $t \leq t_{C}$, the mismatch between $S(t)$ and $S^{e}(t)$ is not problematic; rather, it is an expected and reasonable.

\section{Real-data experiments}\label{sec:sec6}
\subsection{Experiment of the Binarized MNIST for the speciation time}\label{subsec:sec6_1}

So far, our analysis has been based on artificial data generated from the effective model. We now turn to a comparison between the theoretical predictions for $t_{S}$ and $t_{C}$ and the results obtained from training and generating data with an actual discrete diffusion model on real data. As a first step toward validating $t_{S}$, we train one of the most widely used discrete diffusion models, Discrete Denoising Diffusion Models \cite{austin2021structured} (D3PMs), on a dataset besed on the MNIST \cite{lecun1998gradient}. In order to match our binary theory, we use the binary version of the MNIST \cite{salakhutdinov2008quantitative} (hereafter referred to as BinMNIST). We examine the branching behavior of the trajectories of the generated data in the reverse diffusion process by using BinMNIST.

To clearly distinguish the branching of trajectories between classes, we focus on two classes with visually distinct digit shapes, namely labels 1 and 8. The size of the data set is $p_{1} = 6742$ for label 1 and $p_{8} = 5851$ for label 8, corresponding to all training data from labels 1 and 8 in the MNIST dataset. The data dimensionality is $N = 28 \times 28 = 784$. In the generation phase after training, we employ conditional generation based on the label.

In the theoretical analysis, we assumed that the noise level is constant at each time step. In practical diffusion models, however, the noise schedule is typically designed to increase gradually with time.
This requires to replace $\theta^t$ in the theoretical analysis with the cumulative signal $\Theta_t = \prod_{s=1}^t(1-\beta_s)$. Nevertheless, the argument summarized in Eq. (\ref{eq def noise schedule free ts criterion}) is still applicable, which determines the speciation time by the condition 
$1 = [\prod_{s=1}^t(1-\beta_s)]^2 \Lambda$.
Taking the logarithm gives $2\sum_{s=1}^{t_S}\log(1-\beta_s)+\log\Lambda=0$.
In the D3PM experiment on BinMNIST, we use the linear noise schedule described in Appendix \ref{app:appE}. The number of diffusion timesteps is $T=500$, and the noise strength is linearly interpolated from $10^{-4}$ to $0.02$. Thus, we write $\beta_s=a(s-1)+b$, where $a=(0.02-10^{-4})/(500-1)$ and $b=10^{-4}$. Since the noise strength remains small throughout the process, in particular $\beta_T=0.02\ll1$, the approximation $\log(1-\beta_s)\simeq-\beta_s$ is justified. The criterion then reduces to $2\sum_{s=1}^{t_S}\beta_s\simeq\log\Lambda$. Using $\sum_{s=1}^{t_S}\beta_s=\frac{a}{2}t_S(t_S-1)+bt_S$, and denoting by $t_S^{\mathrm{lin}}$ the speciation time for the linear noise schedule, we obtain
\begin{eqnarray}
t_{S}^{\mathrm{lin}}
=
\frac{a-2b+\sqrt{(2b-a)^2+4a\log\Lambda}}{2a},
\label{eq ts linear schedule}
\end{eqnarray}
where we used $t > 0$. For the parameter values used in our experiment, $\Lambda=32.77$, $a=(0.02-10^{-4})/(500-1)$, and $b=10^{-4}$, this gives $t_S^{\mathrm{lin}}\simeq 293.80$. For real data, $\Lambda$ corresponds to the largest eigenvalue of the empirical covariance matrix of the data. In this experiment, more specifically, $\Lambda$ is given by the largest eigenvalue of the empirical covariance matrix computed from all training data of labels 1 and 8.

\begin{figure}[t]
\centering
 \includegraphics[width=9cm]{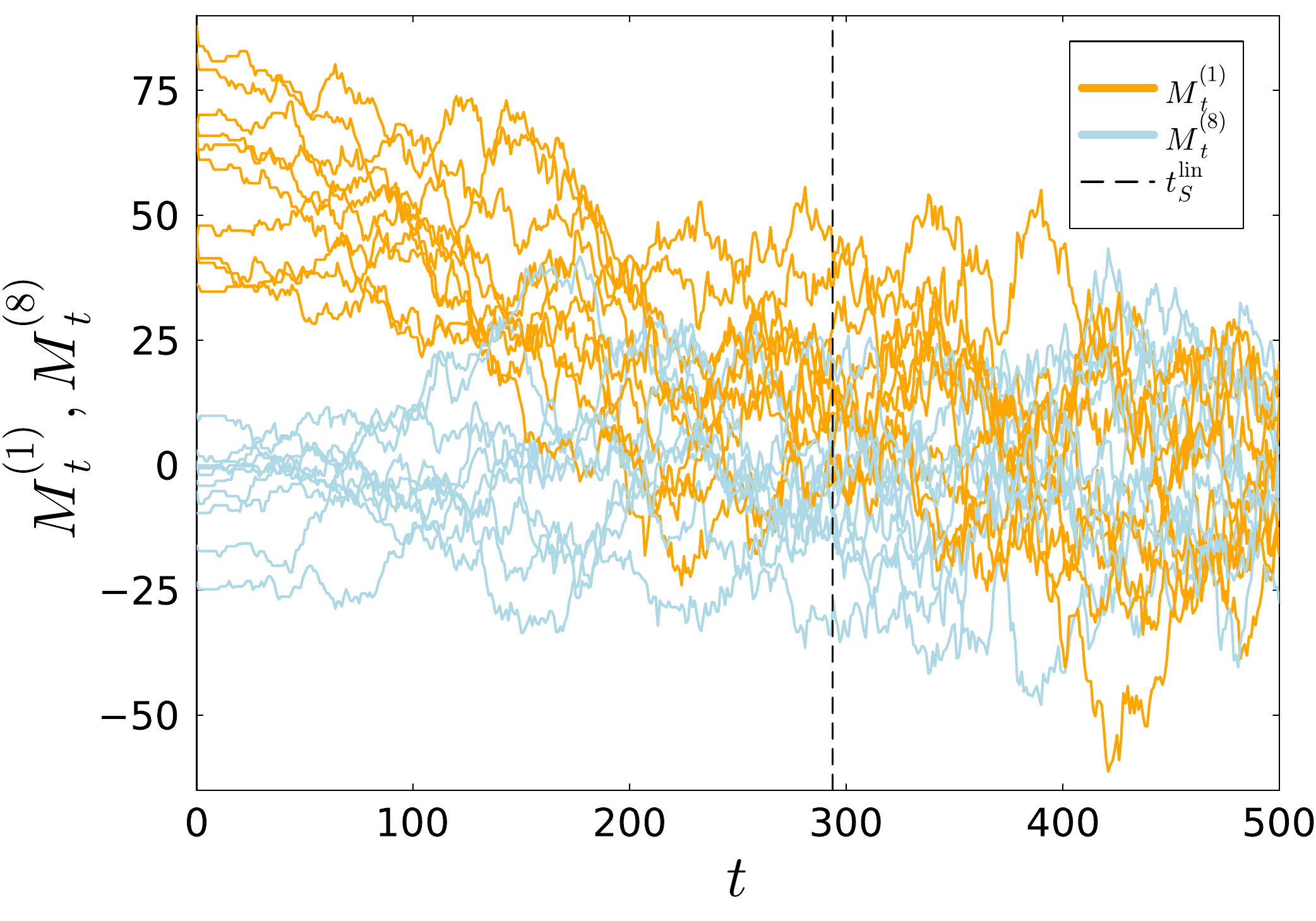}
 \caption{
 Trajectories of generated data in the backward process of D3PM on BinMNIST, together with the theoretical prediction of $t_{S}^{\mathrm{lin}}$ given by Eq. (\ref{eq ts linear schedule}), shown as a dashed vertical line. The orange trajectories represent conditional generation for label 1, $M_{t}^{(1)}$, while the light blue trajectories correspond to conditional generation for label 8, $M_{t}^{(8)}$. Ten trajectories are shown for each label. The result of the value of $t_{S}^{\mathrm{lin}}$ is $t_{S}^{\mathrm{lin}} = 293.80$.
 }
\end{figure}

Let $\bm e_{t}^{(1)}$ and $\bm e_{t}^{(8)}$ denote the $N$-dimensional generated data vectors at each time step of the backward process for labels 1 and 8, respectively. We then compute the inner products between these vectors and the vector whose components are the averages, over all training samples, of the pixel values at each fixed pixel location for the corresponding label (hereafter we call it average vector). However, both the inner product with respect to label 1 and that with respect to label 8 converge to nearly the same value close to 1 at the end of the backward dynamics, making the branching unobservable. Therefore, instead of using $\bm e_{t}^{(1)}$ and $\bm e_{t}^{(8)}$, we use the corresponding relative vectors, $\tilde{\bm e_{t}}^{(1)} = (\bm e_{t}^{(1)} - \bm e_{t}^{(8)})/2$ and $\tilde{\bm e_{t}}^{(8)} = (\bm e_{t}^{(8)} - \bm e_{t}^{(1)})/2$, and compute the following quantities:
\begin{eqnarray}
M_{t}^{(1)} = \bar{\bm x}^{(1)} \cdot \tilde{\bm e_{t}}^{(1)}, \qquad
M_{t}^{(8)} = \bar{\bm x}^{(8)} \cdot \tilde{\bm e}_{t}^{(8)}\label{eq def M_t_8},
\end{eqnarray}
where $\bar{\bm x}^{(1)}$ and $\bar{\bm x}^{(8)}$ are the average vectors of labels 1 and 8, respectively.

Fig. 4 shows the trajectories of generated data in backward process and the linear shcedule version of the theoretical prediction of $t_{S}$ at Eq. (\ref{eq ts linear schedule}). As can be seen from Fig. 4, the theoretical prediction for $t_{S}$ successfully captures, to a good approximation, the timing of the branching between labels. In addition, qualitatively, fluctuations are large in the large regime $t$, while they gradually decrease as $t$ approaches $0$, that is, as the system approaches the end point of the backward process.

As shown in Fig. 4, $M_{t}^{(8)}$ tends to take slightly negative values after speciation. According to its definition in Eq. (\ref{eq def M_t_8}), this means that $\bar{\bm x}^{(8)} \cdot \bm e_{t}^{(8)} < \bar{\bm x}^{(8)} \cdot \bm e_{t}^{(1)}$, which may seem counterintuitive from the viewpoint that samples generated toward label 8 should have a larger overlap with the average vector of label 8. However, considering the large diversity of MNIST images with label 8, it is still possible that, even after speciation, the overlap defined in this way is on average larger for label 1. We therefore do not regard this feature as affecting the conclusion.

To obtain a more reliable validation, we compute the cloning probability in the same manner as in Sec. \ref{subsec:sec5_1}. However, in the present case, since the cloning analysis must be performed with respect to the empirical distribution constructed from the BinMNIST training data, it is necessary to use a method based on the empirical marginal distribution $P_{t}^{e}(\bm x_{t})$, rather than the marginal-distribution-based approach using $P_{t}(\bm x_{t})$. The cloning probability obtained by the latter approach, $\phi_{S}^{e}(t)$, is given as following empirical average.
\begin{eqnarray}
    \phi_{S}^{e}(t) 
     = 
    \sum_{\bm x_{t}} \frac{\left(\eta\sum_{\mu \in C_{1}} e^{F_{t} \bm x_{t} \cdot \bm x^{\mu}}\right)^{2} + \left((1-\eta)\sum_{\mu \in C_{2}} e^{F_{t} \bm x_{t} \cdot \bm x^{\mu}}\right)^{2}}{
    \left(\eta\sum_{\mu \in C_{1}} e^{F_{t} \bm x_{t} \cdot \bm x^{\mu}} + (1-\eta)\sum_{\mu \in C_{2}} e^{F_{t} \bm x_{t} \cdot \bm x^{\mu}}\right)^{2}
    }\nonumber\\
    \hspace{20mm} \times \frac{1}{p} \frac{1}{[2\cosh F_{t}]^{N}} \left( \sum_{\mu = 1}^{p} e^{F_{t} \bm x_{t}\cdot \bm x^{\mu}}\right)\label{eq def empirical cloning prob speciation},
\end{eqnarray}
where $C_{1}$ denotes the set of indices of data points belonging to one class (label), while $C_{2}$ denotes the set of indices of data points belonging to the other class. All other parameters are defined in the same way as in the previous sections. The derivation of Eq. (\ref{eq def empirical cloning prob speciation}) is presented in Appendix \ref{app:appB}. The empirical cloning probability $\phi_{S}^{e}(t)$ has the same interpretation as that described in Sec. \ref{sec:sec5}: it is the probability that two trajectories which share the same configuration at time $t$ belong to the same class in the training data at time $0$.

{\color{purple} }

\begin{figure}[t]
\centering
 \includegraphics[width=9cm]{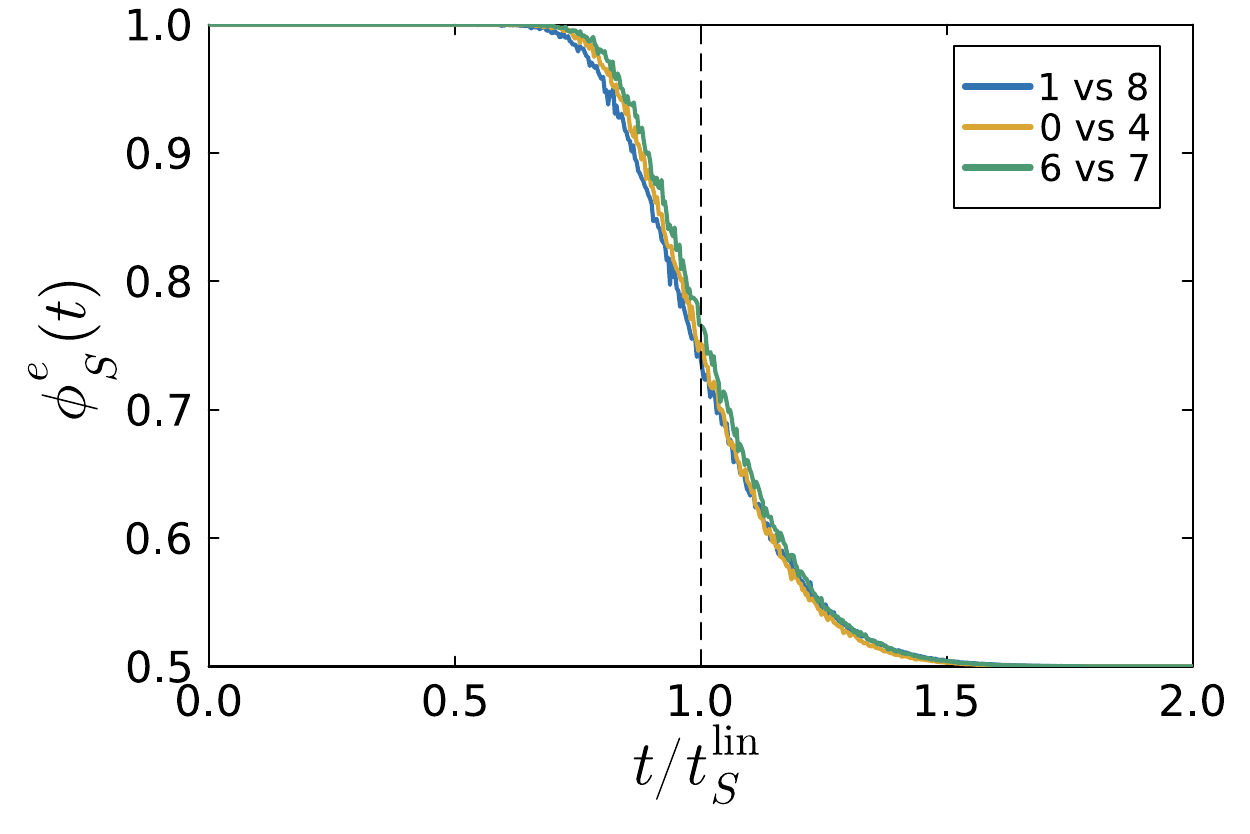}
 \caption{Empirical cloning probability $\phi_{S}^{e}(t)$ for different pair of labels of BinMNIST. The horizontal axis represents time rescaled by $t_{S}^{\mathrm{lin}}$, we use the expression given in Eq. {(\ref{eq ts linear schedule})}. The blue, yellow, and green curves correspond to the label pairs (1, 8), (0, 4), and (6, 7), respectively. For each pair, to let $\eta = 0.5$, $\phi_{S}^{e}(t)$ is computed using the same number of training samples, matched to the smaller class in the pair. We set $p = 11702$ for labels 1 and 8, $p = 11684$ for labels 0 and 4, and $p = 11836$ for labels 6 and 7.}
\end{figure}

Fig. 5 shows the $\phi_{S}^{e}(t)$ for different pair of labels of BinMNIST. The horizontal axis represents time rescaled by $t_{S}^{\mathrm{lin}}$, as in Fig. 2. In addition, the mixing proportion is fixed to $\eta = 0.5$ for all label pairs. Specifically, for each pair, $\phi_{S}^{e}(t)$ is computed using the same number of training samples, matched to the smaller class in the pair (namely, $p = 11702$ for labels 1 and 8, $p = 11684$ for labels 0 and 4, and $p = 11836$ for labels 6 and 7). As can be seen from Fig. 5, the empirical cloning probabilities $\phi_{S}^{e}(t)$ for the three label pairs intersect at $t = t_{S}$. Thus, following the same argument as in Sec. \ref{subsec:sec5_1}, the theoretical prediction for $t_{S}$ is found to be consistent with the results for BinMNIST.
    
\subsection{Experiment of the Binarized MovieLens Tag Genom for the collapse time}\label{subsec:sec6_2}

For real-world datasets with strong correlations among individual data points, such as MNIST, it is difficult to observe the collapse itself. We therefore consider uncorrelated discrete data and use a binarized version of the Tag Genome from the MovieLens dataset (hereafter referred to as BinMLTG), which consists of data such as movie ratings \cite{movielens}. The MovieLens Tag Genome (MLTG) quantifies, for each assigned ``tag," the degree of relevance between a movie and that tag as a continuous value between 0 and 1 \cite{vig2012taggenome}. The tags refer to descriptive attributes of movies, such as ``catastrophe" or ``romantic." Since the ordering of tags carries no semantic meaning, the MovieLens Tag Genome data can be regarded as uncorrelated across variables. Binarization is performed independently for each tag by assigning values less than 0.5 to 0 and values greater than or equal to 0.5 to 1. For the analysis, in order to match the effective model proposed in this work, the value 0 is further mapped to $-1$. The MovieLens Tag Genome has been used as a dataset for developing recommendation systems based on generative models, often referred to as generative recommendation \cite{vothanh2018generation}.

\begin{figure*}[t]
\centering

% ===== upper row =====
\begin{minipage}{0.48\linewidth}
    \centering
    \includegraphics[width=\linewidth]{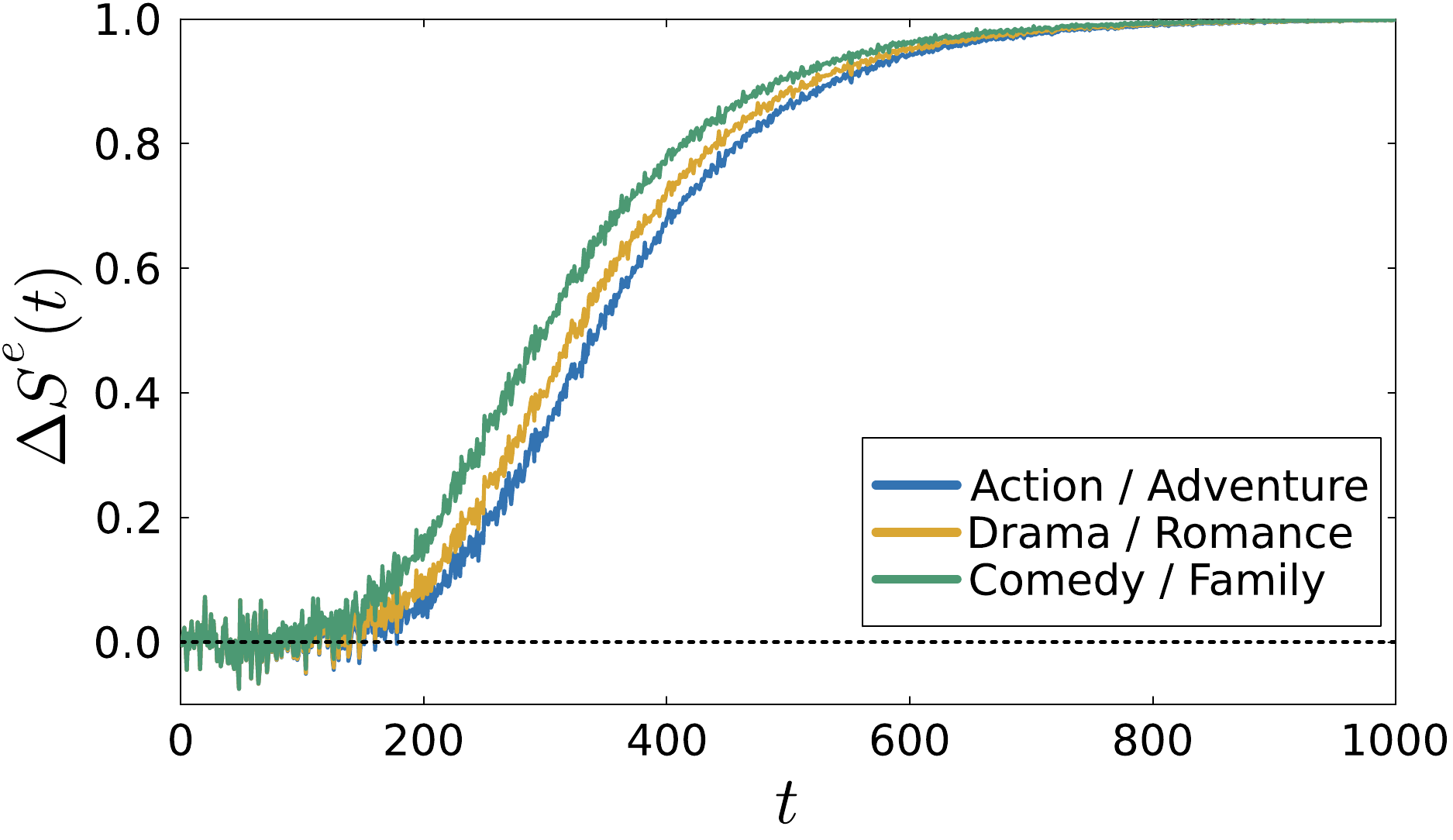}\\
    (a) $\Delta S^{e}(t)$
\end{minipage}
\hfill
\begin{minipage}{0.48\linewidth}
    \centering
    \includegraphics[width=\linewidth]{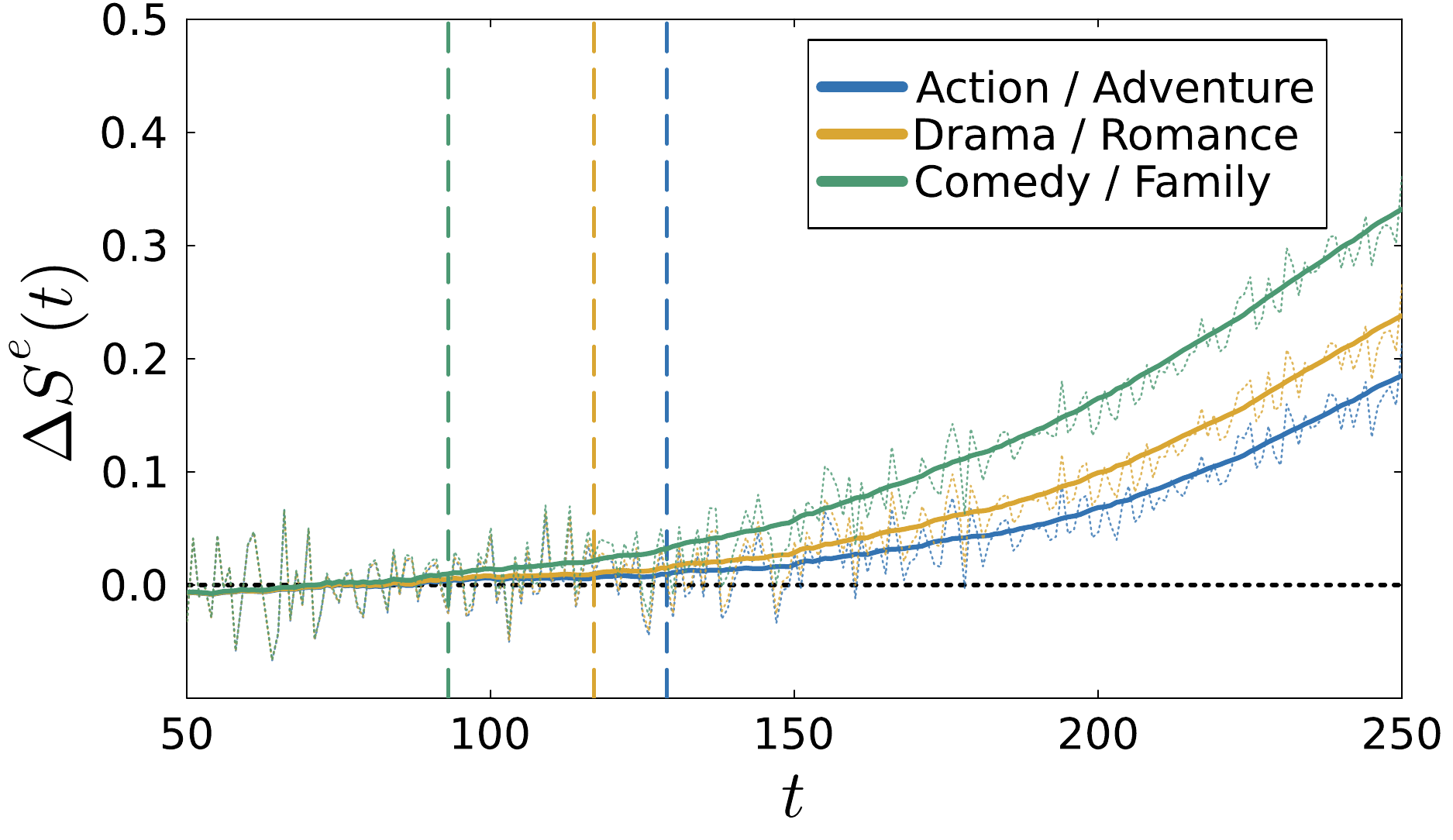}\\
    (b) $\Delta S^{e}(t)$ near $t_C^{\mathrm{emp}}$
\end{minipage}

\vspace{3mm}

% ===== lower row =====
\begin{minipage}{0.65\linewidth}
    \centering
    \includegraphics[width=\linewidth]{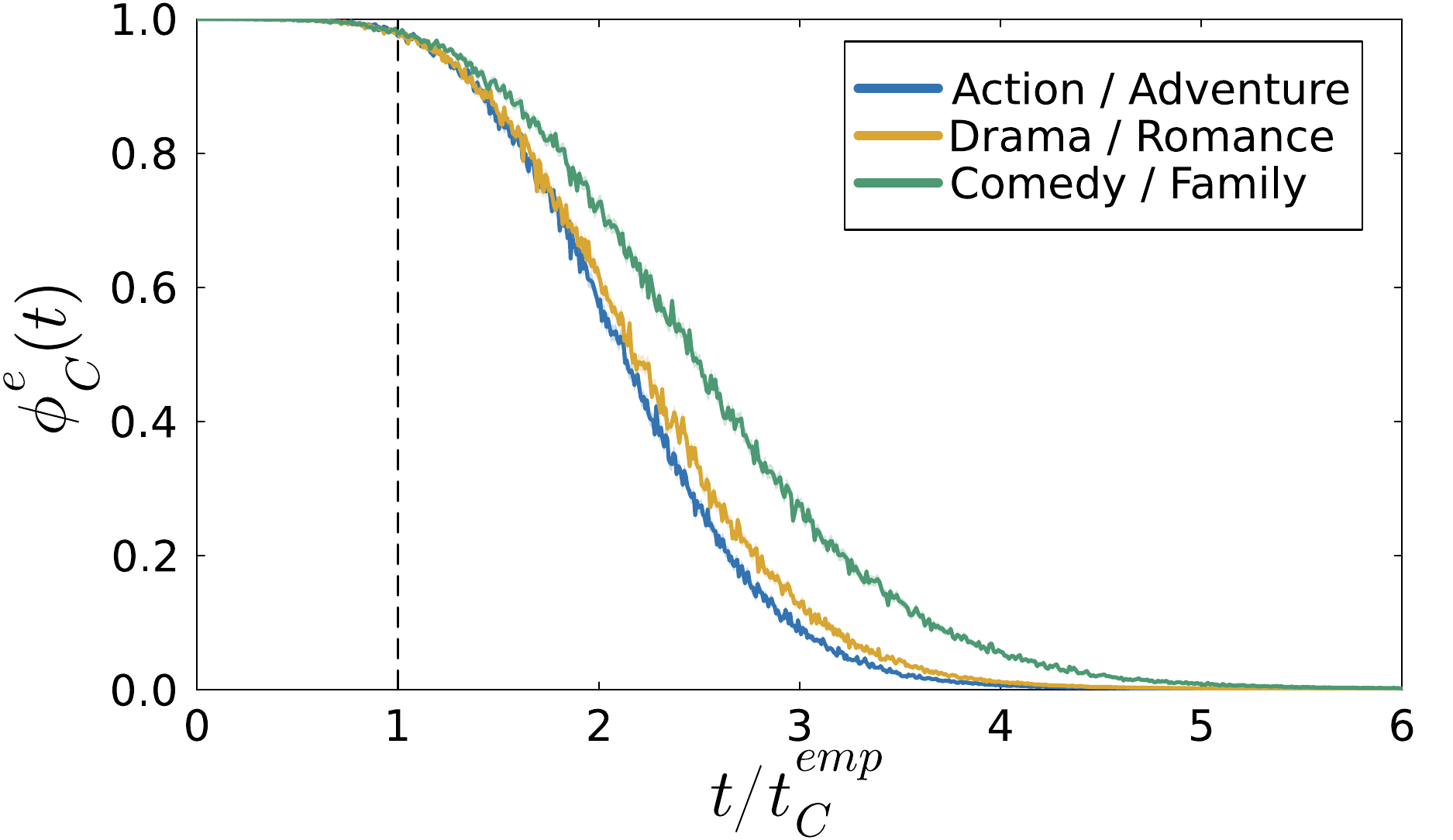}\\
    (c) $\phi_{C}^{e}(t)$
\end{minipage}

\caption{
(a): Backward dynamics of the entropy difference $\Delta S^{e}(t)$ for the three movie groups. The sample size at each time step ($n_{sample}$ in Eq. (\ref{eq def S_e_t})) is set to $10000$.
(b): Enlarged view of the region around $\Delta S^{e}(t)=0$ extracted from panel (a). Solid lines show a centered moving average over 50 time steps, while thin dotted lines indicate the corresponding raw trajectories. Vertical dashed lines denote the empirical collapse times $t_{C}^{emp}$, defined as the times at which the moving-averaged curves first reach zero within a tolerance of $0.01$.
(c): Cloning probability $\phi_{C}^{e}(t/t_{C}^{emp})$ plotted as a function of the rescaled time for each genre, where $t_{C}^{emp}$ is determined from panel (b). The sample size used at each time step for the computation of $\phi_{C}^{e}(t)$ (denoted by $\pi$ in Eq. (\ref{eq phi_C_e empirical ave}) of Appendix \ref{app:appB}) is $1000$.
}
\label{fig:comparison}
\end{figure*}

We here analyze BinMLTG data associated with the following three genres: \textit{Action/Adventure, Drama/Romance, and Comedy/Family}.
In the MovieLens dataset, genre attributes are assigned to each movie. However, since many movies are labeled with multiple genres, it is difficult to construct mutually exclusive datasets based on a single genre.
Therefore, we construct three mutually exclusive datasets, each consisting of 1000 movies that share two genres with substantial overlap, as described above.
The number of tags is fixed to $N = 1128$ for all movies, which corresponds to the default number of tags in the MLTG representation.

From the results in Sec. \ref{subsec:sec5_1}, we find that the entropy difference $\Delta S^{e}(t)$ constructed from the empirical marginal distribution $P_{t}^{e}(\bm x_{t} | \mathcal{D})$ yields accurate values once finite-size effects are properly taken into account.
Therefore, in the following analysis, we determine the empirical collapse time $t_{C}^{emp}$ for each of the three movie groups, defined by the condition $\Delta S^{e}(t_{C}^{emp}) = 0$, and examine the collapse behavior by observing the cloning probability rescaled as $t / t_{C}^{emp}$.
The cloning probability considered here is defined in accordance with the nature of the collapse: it is the probability that two generated data points which share exactly the same configuration at time $t$ reach the same training data point at $t = 0$. It is given by,
\begin{eqnarray}
    \phi_{C}^{e}(t)
    &=&
    \sum_{\bm x_{t}} 
    \sum_{\mu = 1}^{p} \left(  \frac{e^{F_{t} \bm x_{t} \cdot \bm x^{\mu}}}{\sum_{\nu = 1}^{p}e^{F_{t} \bm x_{t} \cdot \bm x^{\nu}}} \right)^{2} \frac{1}{p} \sum_{\sigma = 1}^{p}\frac{e^{F_{t} \bm x_{t} \cdot \bm x^{\sigma}}}{[2\cosh F_{t}]^{N}}\label{eq phi_C_e main text}.
\end{eqnarray}
The derivation of Eq. (\ref{eq phi_C_e main text}) follows directly from the cloning probability for sharing the same class presented in Sec. \ref{subsec:sec5_1} by specifying the class as an individual data point (see Appendix \ref{app:appB}).

Fig. 6 shows those results. Fig. 6 (a) shows the backward dynamics of $\Delta S^{e}(t)$ for the above three movie groups. The behavior of the curves of $\Delta S^{e}(t)$ is very similar to the toy data version shown in Fig. 3 (b). Figure 6(b) shows a magnified view of the region around $\Delta S^{e}(t)=0$ extracted from Fig. 6 (a). The solid lines represent a moving average over 50 time steps, while the superimposed thin dotted lines correspond to the raw data, which are identical to those shown in Fig. 6 (a). The vertical dashed lines indicate the times at which the moving-averaged curves first reach zero. As can be seen from this panel, determining $t_{C}^{emp}$ directly from the raw dynamics is difficult due to strong temporal fluctuations; for this reason, a moving average is employed to estimate the collapse time.
Fig. 6 (c) displays the cloning probability $\phi_{C}^{e}(t/t_{C}^{emp})$ for each genre, where $t_{C}^{emp}$ is obtained from Fig. 6(b). From Fig. 6(c), we observe that the cloning probability $\phi_{C}^{e}(t)$ takes nearly the same value at $t_{C}^{emp}$ across all genres. Within the framework of finite-size scaling, this result confirms that the criterion $\Delta S^{e}(t)=0$ provides a valid definition of the collapse time even for real-world data. Furthermore, the genre-dependent differences in the temporal evolution of $\Delta S^{e}(t)$ and $\phi_{C}^{e}(t)$ may reflect differences in the diversity of movies across genres.

\section{Summary}

We addressed how the dynamical regimes in the backward process of discrete generative diffusion is in the perspective of statistical physics of disordered systems.
We proposed an simple effective model of the discrete generative diffusion in which the data consist of non-interacting two-component Ising variables (Sec. \ref{sec:sec3}).
Using a high-temperature expansion, we identified the speciation transition with a second-order transition and derived the general criterion for the speciation time, Eq. (\ref{eq def noise schedule free ts criterion}). From this criterion, we derived Eq. (\ref{eq ts}) as the speciation time for constant noise schedules, and Eq. (\ref{eq ts correspond to continuous}) for general time-dependent noise schedules. We further showed that the latter result is consistent, at the level of the leading scaling, with previous studies for continuous variables and for discrete variables under more general conditions (Sec. \ref{subsec:sec4_1}).
We also derived the analytic condition of the collapse time $t_{C}$ that Eq. (\ref{eq st main text}) equals zero as the condensation transition of the disordered systems by the Random Energy Model (Sec. \ref{subsec:sec4_2}). 
In the collapse analysis, we show that the partition function is correctly evaluated when the energy distribution is conditioned on the data generated at each time step.

Through numerical and real-data experiments, we find that the theoretical predictions of the speciation time and the collapse time can capture the bifurcation point of backward trajectories and the general criterion for collapse, respectively. This result is further validated by the cloning analysis (Sec. \ref{sec:sec5} and \ref{sec:sec6}).

These results demonstrate that discrete diffusion models exhibit the same three dynamical phases as continuous-variable diffusion models, under the same criteria for those phase boundaries.

For future work, it will be important to extend the present theory to settings with more classes and categories, as well as to cases involving interactions among variables, such as in graph data. Such extensions would provide a deeper understanding of the dynamical regimes of more practically relevant discrete diffusion models.

\section*{Acknowledgement}
\addcontentsline{toc}{section}{Acknowledgement}

The authors acknowledge financial support from JSPS KAKENHI Grant Numbers 25K21296 and 23K19996 (Tomoei Takahashi).
We thank Koki Okajima for assistance with the sampling method for the backward process.
We also thank Beatrice Achilli, Tony Bonnaire, Enrico Ventura, and Carlo Lucibello for illuminating discussions and helpful comments.
This work was also supported by JSPS KAKENHI Grant
Numbers 22H05117 and 23K16960, and JST ACT-X Grant Number
JPMJAX24CG (Takashi Takahashi), and JSPS KAKENHI Grant
Number 22H05117 (Y.K.).

\setcounter{section}{0}
\renewcommand{\thesection}{\Alph{section}}

\refstepcounter{section}
\section*{Appendix \thesection. The sampling in the reverse process}
\addcontentsline{toc}{section}{Appendix \thesection. The sampling in the reverse process}
\label{app:appA}
\renewcommand{\theequation}{\thesection.\arabic{equation}}
\setcounter{equation}{0}

Through the Bayes theorem, the reverse process $P(\bm x_{t-1} | \bm x_{t})$ becomes
\begin{eqnarray}
 P(\bm x_{t-1} | \bm x_{t}) = \frac{P(\bm x_{t} | \bm x_{t-1}) P_{t-1}(\bm x_{t-1})}{P_{t}(\bm x_{t})}
     \propto  P_{t-1}(\bm x_{t-1}) P(\bm x_{t} | \bm x_{t-1})\label{eq backward bayes}\\
		= \eta \prod_{i = 1}^{N}  \frac{1 + \theta^{t-1} m x_{t-1,i}}{2}  +(1-\eta)\prod_{i = 1}^{N}  \frac{1 - \theta^{t-1} m x_{t-1,i}}{2}\nonumber\\
        \times \prod_{i = 1}^{N} \frac{1+\theta x_{ti} x_{t-1,i}}{2}\label{eq backward bayes line 2}.
\end{eqnarray}
Sampling $\bm x_{t-1}$ from Eq. (\ref{eq backward bayes line 2}) becomes computationally intractable when $N$ is large. Therefore, by applying several calculations to Eq. (\ref{eq backward bayes line 2}), as described below, we obtain a reformulated expression that enables highly efficient sampling.

To derive the sampling method, we define the following gauge transformation
\begin{eqnarray}
    s_{ti} = x_{ti}x_{t-1,i}
\end{eqnarray}
Let $\bm s_{t}$ denotes the following $N$-dimensional vector: $\bm s_{t} = (s_{t1}, s_{t2}, \cdots, s_{tN})$. Then, 
\begin{eqnarray}
    P(\bm x_{t} | \bm x_{t-1}) &=& \prod_{i = 1}^{N}\frac{1+\theta s_{ti}}{2} =: P_{t}(\bm s_{t}),\\
    P_{t-1}(\bm x_{t-1}) &=& \eta\prod_{i = 1}^{N}\frac{1+m\theta^{t-1}x_{ti}s_{ti}}{2} + (1 - \eta)\prod_{i = 1}^{N}\frac{1-m\theta^{t-1}x_{ti}s_{ti}}{2}\label{eq backward Pt-1}\\
    &=:& P(\bm x_{t} | \bm s_{t})\label{eq backward Pxt given st}
\end{eqnarray}
From here we focus on the two new distributions $P_{t}(\bm s_{t})$ and $P(\bm x_{t} | \bm s_{t})$ to derive the algorithm of the sampling in the reverse process. The probability distribution of the gauge variable $P_{t}(\bm s_{t})$ can be regarded as a binomial distribution with respect to the number of components $k = (N - \sum_{i = 1}^{N} s_{ti})/2$, $P(k)$, where $s_{ti} = -1$ indicates that the spin flips during the transition from time $t$ to $t-1$ eliminating the degree of freedom of the configuration of $\bm s_{t}$. Since the value of the probability $P_{t}(\bm s_{t})$ does not depend on the specific configuration once $k$ is fixed, considering the probability distribution of $k$ alone does not lead to any loss of generality. The probability distribution function of $P(k)$ is given by,
\begin{eqnarray}
    P(k) = {N \choose k}\left(\frac{\beta}{2} \right)^{k}\left(1- \frac{\beta}{2}\right)^{N-k},
\end{eqnarray}
where we used $\theta = 1 - \beta$. Here we assume $\beta = \frac{\gamma}{N}$ where $\gamma = \mathcal{O}(1)$. Then, in the limit $N \rightarrow \infty$, the binomial distribution can be approximated by a Poisson distribution, and thus $P(k)$ is given as follows.
\begin{eqnarray}
    P(k) = \frac{(\gamma/2)^{k}}{k!}e^{-\gamma/2}\label{eq backward okajima prior}.
\end{eqnarray}
Sampling from this ``prior" Eq. (\ref{eq backward okajima prior}), determines the number of spins that flip from $t$ to $t - 1$ for any $t$. However, as is clear from Eq. (\ref{eq backward Pt-1}), the locations of the spins that flip at time $t-1$ depend on the signs of $x_{ti}$. Therefore, in the transition from time $t$ to $t-1$, it is necessary to construct a probability distribution for how many spins with $x_{ti} = 1$ flip to $-1$ (or, equivalently, how many spins with $x_{ti} = -1$ flip to $+1$). This can be achieved by rewriting the ``likelihood" $P(\bm x_{t} | \bm s_{t})$ as follows.

Here, for an arbitrary time $t > 1$, we define $A_{t}$ as the set of indices satisfying $x_{ti} = 1$, and $\bar{A_{t}}$ as the set of indices satisfying $x_{ti} = -1$. Then, each term of $P(\bm x_{t} | \bm s_{t})$ can be expressed as a product of four distinct factors, as shown below.
\begin{eqnarray}
    P(\bm x_{t}|\bm s_{t}) = \eta \left( \prod_{i \in A_{t} \cap A_{t-1}}  \frac{1 + m\theta^{t-1}}{2}\right)\left( \prod_{i \in A_{t} \cap \bar{A}_{t-1}}  \frac{1 - m\theta^{t-1}}{2}\right)
    \nonumber\\
    \times
    \left( \prod_{i \in \bar{A}_{t} \cap A_{t-1}}  \frac{1 - m\theta^{t-1}}{2}\right)\left( \prod_{i \in \bar{A}_{t} \cap \bar{A}_{t-1}}  \frac{1 + m\theta^{t-1}}{2}\right)
    \nonumber\\
    +
    (1-\eta) \left( \prod_{i \in A_{t} \cap A_{t-1}}  \frac{1 - m\theta^{t-1}}{2}\right)\left( \prod_{i \in A_{t} \cap \bar{A}_{t-1}}  \frac{1 + m\theta^{t-1}}{2}\right)
    \nonumber\\
    \times
    \left( \prod_{i \in \bar{A}_{t} \cap A_{t-1}}  \frac{1 + m\theta^{t-1}}{2}\right)\left( \prod_{i \in \bar{A}_{t} \cap \bar{A}_{t-1}}  \frac{1 - m\theta^{t-1}}{2}\right)\label{eq backward four product}
\end{eqnarray}
Furthermore, we define
$k_{1} := |A_{t} \cap \bar{A}_{t-1} |$ and
$k_{2} := | \bar{A}_{t} \cap A_{t-1} |$, where $|\cdot|$ denotes the size of given set.
Then, up to the degeneracy associated with the choice of spin positions that flip, Eq. (\ref{eq backward four product}) can be identified to following mixture distribution $P(k_{1}, k_{2})$ of joint binomial distributions in $k_{1}$ and $k_{2}$.
\begin{eqnarray}
    P(k_{1}, k_{2}) = \eta{|A_{t}| \choose k_{1}} \left( \frac{1 + m\theta^{t-1}}{2} \right)^{|A_{t}| - k_{1}}\left( \frac{1 - m\theta^{t-1}}{2} \right)^{k_{1}} \nonumber\\
    \times
    {|\bar{A}_{t}| \choose k_{2}} \left( \frac{1 - m\theta^{t-1}}{2} \right)^{|\bar{A}_{t}| - k_{2}}\left( \frac{1 + m\theta^{t-1}}{2} \right)^{k_{2}}\nonumber\\
    +
    (1-\eta) {|A_{t}| \choose k_{1}} \left( \frac{1 - m\theta^{t-1}}{2} \right)^{|A_{t}| - k_{1}}\left( \frac{1 + m\theta^{t-1}}{2} \right)^{k_{1}} \nonumber\\
    \times
    {|\bar{A}_{t}| \choose k_{2}} \left( \frac{1 + m\theta^{t-1}}{2} \right)^{|\bar{A}_{t}| - k_{2}}\left( \frac{1 +- m\theta^{t-1}}{2} \right)^{k_{2}}
\end{eqnarray}
However, due to the constraint $k_{2} = k - k_{1}$, once $k_{1}$ is sampled, the value of $k$ is already fixed by the prior distribution, Eq. (\ref{eq backward okajima prior}). Consequently, $k_{2} = k - k_{1}$ holds with probability one. Therefore, Eq. (\ref{eq backward four product}) can be written as a mixture binomial distribution depending only on $k_{1}$ as follows:
\begin{eqnarray}
\hspace{-15mm}
    P(k_{1}) = \eta \hspace{1mm}
Bin\!\left(
k_{1} \Big| |A_{t}|, \frac{1 - m\theta^{t-1}}{2}
\right)
+ (1-\eta) \hspace{1mm}
Bin\!\left(
k_{1} \Big| |A_{t}|, \frac{1 + m\theta^{t-1}}{2}
\right)\label{eq backward Pk1}
\end{eqnarray}
After observing the generated data $\bm x_{t}$ at time $t$, a single sample is drawn from the probability distribution in Eq. (\ref{eq backward Pk1}). This sample specifies how many of the $k$ flipping spins are assigned to $k_{1}$. The remaining spins are then assigned to $k_{2}$. 
Because the positions of these spins are irrelevant, once a value of $k_{1}$ is sampled, one can construct $\bm x_{t-1}$ by randomly selecting $k_{1}$ spins from $A_{t}$ and flipping them, and independently selecting $k_{2} = k - k_{1}$ spins from $\bar{A}_{t}$ and flipping them.

In this way, at each time step, the sampling procedure requires only a single draw from the Poisson distribution in Eq. (\ref{eq backward okajima prior}), a single draw from the mixture binomial distribution in Eq. (\ref{eq backward Pk1}), and three uniform samplings to choose the spin positions corresponding to $k$, $k_{1}$, and $k_{2}$, respectively, resulting in a substantial improvement in sampling efficiency. This sampling scheme yields exact samples from the original probability distribution, Eq. (\ref{eq backward bayes}), in the limit $N \to \infty$.

\refstepcounter{section}
\section*{Appendix \thesection. The derivation of the partition function and the microcanonical entropy density of the REM for the collapse time}
\addcontentsline{toc}{section}{Appendix \thesection. The derivation of the partition function and the microcanonical entropy density of the REM for the collapse time}
\label{app:appB_rev}

We define the following ``energy".
\begin{eqnarray}
    \varepsilon = \frac{1}{N} \sum_{i = 1}^{N}x_{ti}x_{i}^{\mu}.
\end{eqnarray}
To determine the probability distribution of the energy $\varepsilon$, we consider the probability distribution of the overlap $y_{ti} = x_{ti} x_{i}^{\mu}$. We define the vector $\bm y_{t}$ by $\bm y_{t} = (y_{t1}, \ldots, y_{tN})^{\top}$.
The crucial point is that the fluctuations that we focus on concern the event of how similar $\bm x_{t}$, obtained at each time $t$ of the backward process, is to an arbitrary data point $\bm x^{\mu}$.
Therefore, the probability relevant for $\bm y_{t}$ should be taken as the conditionalprobability with $\bm x_{t}$ fixed, namely, $P(\bm y_{t} | \bm x_{t})$. This indicates that the conditional distribution of $\varepsilon$ given $\bm{x}_t$ is evaluated as
\begin{eqnarray}
    P(\varepsilon | \bm x_{t}) &=& \sum_{\bm y_{t}} P(\bm y_{t} | \bm x_{t})\delta(\bm x_{t}\cdot \bm x^{\mu} - N\varepsilon)\\
    &=&
    \sum_{\bm y_{t}} P(\bm y_{t} | \bm x_{t})\delta \left(\sum_{i = 1}^{N}y_{ti} - N\varepsilon \right).
\end{eqnarray}

The conditional distribution 
$P(\bm{y}_t| \bm{x}_t)$ is assessed as follows.
Clearly, if $x_{ti} = 1$, then $y_{ti} = x_{i}^{\mu}$, whereas if $x_{ti} = -1$, we have $y_{ti} = -x_{i}^{\mu}$. Hence, the distribution of $y_{ti}$ can be expressed in terms of conditional probabilities with respect to $x_{ti}$, $P_{ti}(y_{ti} | x_{ti} = 1) = P_{0i}^{+}(x_{i}^{\mu})$ and $P_{ti}(y_{ti} | x_{ti} = -1) = P_{0i}^{+}(-x_{i}^{\mu})$, where $P_{0i}^{+}$ denotes the single-site factor of $P^+$, given in Eq. (\ref{eq def P_plus and P_minus}). Near the collapse time, the system has already branched into either the $+$ or the $-$ class. Since we focus here on the $+$ class, the mixing proportion is set to $\eta = 1$. Accordingly, we have $P_{0i}^{+}(x_{i}^{\mu}) = \frac{1 + m x_{i}^{\mu}}{2}$. Enumerating all possible cases, we obtain the following.
\begin{eqnarray}
    P_{ti}(y_{ti} = 1 | x_{ti} = 1) &=& \frac{1 + m}{2}\\
    P_{ti}(y_{ti} = -1 | x_{ti} = 1) &=& \frac{1 - m}{2}\\
    P_{ti}(y_{ti} = 1 | x_{ti} = -1) &=& \frac{1 - m}{2}\\
    P_{ti}(y_{ti} = -1 | x_{ti} = -1) &=& \frac{1 + m}{2}.
\end{eqnarray} 
Here, let $K$ denote the number of indices $i = 1,\ldots,N$ for which $y_{ti} = 1$, 
let $k$ denote the number of indices $i$ such that $x_{ti} = 1$ and $y_{ti} = 1$, 
and let $N_{+}^{t}$ denotes the number of indices $i = 1,\ldots,N$ for which $x_{ti} = 1$.
From the above considerations, $P(\bm y_{t} | \bm x_{t})$ can be written as a product of independent probabilities for each index $i$. %and any configuration is fully characterized solely by $K$ and $k$.
For the same reason, the dependence on $\bm x_{t}$ is also determined only by $N_{+}^{t}$.
Therefore, $P(\bm y_{t} | \bm x_{t})$ can be expressed as a convolution of the following two binomial distributions:
\begin{eqnarray}
    P(\bm y_{t} | \bm x_{t}) &=& 
    P(\bm y_{t} | N_{+}^{t})\\ &=& 
    %{N_{+}^{t} \choose k}
    \left(\frac{1 + m}{2}\right)^{k}\left(\frac{1 - m}{2}\right)^{N_{+}^{t} - k}
    %{N - N_{+}^{t} \choose K - k}
    \left(\frac{1 - m}{2}\right)^{K - k}\left(\frac{1 + m}{2}\right)^{N - N_{+}^{t} - (K - k)}.
\end{eqnarray}

The number of up spins $N_{+}^{t}$ is indeed a random variable. However, for $N \gg 1$, we may assume that it takes the deterministic value $N_{+}^{t} = \frac{N(1 + m\theta^{t})}{2}$ at any time $t$, neglecting probabilistic fluctuations. In other words, we are here assuming that, with respect to the fluctuations of $\bm x_{t}$, both $P(\bm y_{t} | \bm x_{t})$ and $P(\varepsilon | \bm x_{t})$ satisfy self-averaging property.
Such self-averaging property is expected to hold sufficiently well in the limit $N \rightarrow \infty$ because of the law of large numbers.

Using the constraints $0 \leq k \leq N_{+}^{t}$ and $0 \leq K-k \leq N - N_{+}^{t}$, and rewriting them via $K = \frac{N(1 + \varepsilon)}{2}$, we obtain the following expression for $P(\varepsilon | \bm x_{t})$.
\begin{eqnarray}
    \hspace{-20mm}P(\varepsilon | \bm x_{t}) &=& 
    \sum_{\bm y_{t}}\left(\frac{1 + m}{2}\right)^{k}\left(\frac{1 - m}{2}\right)^{N_{+}^{t} - k}\nonumber \\
    &\times&
    %{N - N_{+}^{t} \choose K - k}
    \left(\frac{1 - m}{2}\right)^{K - k}\left(\frac{1 + m}{2}\right)^{N - N_{+}^{t} - (K - k)}\delta \left( K - \frac{N(1 + \varepsilon)}{2}\right)\\
    &=&
    \sum_{K = 0}^{N}
    \sum_{k = \mathrm{max}(0, K - N + N_{+}^{t})}^{\mathrm{min}(K, N_{+}^{t})
    }
    {N_{+}^{t} \choose k}
    \left(\frac{1 + m}{2}\right)^{k}\left(\frac{1 - m}{2}\right)^{N_{+}^{t} - k}\nonumber\\
    &\times&
    {N - N_{+}^{t} \choose K - k}
    %{N - N_{+}^{t} \choose K - k}
    \left(\frac{1 - m}{2}\right)^{K - k}\left(\frac{1 + m}{2}\right)^{N - N_{+}^{t} - (K - k)}
     \delta \left( K - \frac{N(1 + \varepsilon)}{2}\right).
\end{eqnarray}
Then, we introduce new order parameters defined by
\begin{eqnarray}
u = \frac{k}{N_{+}^{t}}\label{eq def u}\\ 
v = \frac{K-k}{N - N_{+}^{t}}\label{eq def v}.
\end{eqnarray}
In addition, from the trivial constraint $K = K_{+} + K_{-}$ (with $K_{+} = k$ and $K_{-} = K - k$), the following relation holds for $\varepsilon$, $u$, and $v$:
\begin{eqnarray}
    \varepsilon(u, v) = (1 + m\theta^{t})u + (1-m\theta^{t})v - 1\label{eq constraint varepsilon}.
\end{eqnarray}
%where we used $K = \frac{N(1 + \varepsilon)}{2}$ and $N_{+}^{t} = \frac{N(1 + m\theta^{t})}{2}$. 
Using $u$, $v$ and the constraint Eqs. (\ref{eq def u}) and (\ref{eq def v}), $P(\varepsilon | \bm x_{t})$ is further rewritten in the limit of $N \rightarrow \infty$ as
\begin{eqnarray}
    P(\varepsilon | \bm x_{t}) = \int_{0}^{1}\int_{0}^{1}dudv P(u,v | N_{+}^{t})\delta(\varepsilon - \varepsilon(u,v))\label{eq def P_varepsilon},
\end{eqnarray}
where $P(u,v | N_{+}^{t})$ is given by
\begin{eqnarray}
    P(u,v | N_{+}^{t}) = {N_{+}^{t} \choose N_{+}^{t}u}\left( \frac{1 + m}{2}\right)^{N_{+}^{t}u} \left( \frac{1 - m}{2}\right)^{N_{+}^{t}(1-u)}\nonumber\\
    \times
    {N - N_{+}^{t} \choose (N - N_{+}^{t})v}\left( \frac{1 - m}{2}\right)^{(N - N_{+}^{t})v} \left( \frac{1 + m}{2}\right)^{(N - N_{+}^{t})(1-v)}\label{eq P_uv}. 
\end{eqnarray}
Using the large deviation $P(u,v | N_{+}^{t}) = e^{-N I (u,v)}$, where $I(u,v)$ is the rate function, we obtain
\begin{eqnarray}
    P(\varepsilon | \bm x_{t}) = \int_{0}^{1}\int_{0}^{1}dudv\hspace{1mm} e^{-NI(u,v)}\delta(\varepsilon - \varepsilon(u,v)).
\end{eqnarray}

From here, we proceed to the calculation of the partition function $Z_{2...p_{+}}$ using the expression for $P(\varepsilon | \bm x_{t})$ given in Eq. (\ref{eq def P_varepsilon}). The number of states (data points) with energy $\varepsilon$ is $p_{+} P(\varepsilon | \bm x_{t})$. Hence, by the difinition of the partition function, we have
\begin{eqnarray}
    Z_{2...p_{+}} &=& p_{+}\int_{-1}^{1}d \varepsilon \left(\int_{0}^{1}\int_{0}^{1}dudv\hspace{1mm} e^{-NI(u,v)}\delta(\varepsilon - \varepsilon(u,v))\right) e^{NF_{t}\varepsilon}\\
    &=&
    p_{+} \int_{0}^{1} \int_{0}^{1}dudv \hspace{1mm} e^{-N[I(u,v) - F_{t}\varepsilon(u,v)]}\label{eq def Z_plus after delta func integral}\\
    &=&
    \int_{0}^{1}\int_{0}^{1}dudv \hspace{1mm} e^{N[\alpha - I(u,v) + F_{t}\varepsilon(u,v)]}\label{eq def Z_plus after alpha into integral}\\
    &=&
    e^{N[\alpha - I(u^{*},v^{*}) + F_{t}\varepsilon(u^{*},v^{*})]}\label{eq def Z_plus after saddle point method}.
\end{eqnarray}
From Eq. (\ref{eq def Z_plus after delta func integral}) to (\ref{eq def Z_plus after alpha into integral}), we used $\alpha = \frac{\log p}{N}$. Strictly, this quantity is given by $\frac{\log p_{+}}{N}$, however, in the limit $N \to \infty$, we use the fact that the contribution of $\log 2$ becomes negligible. From Eq. (\ref{eq def Z_plus after alpha into integral}) to (\ref{eq def Z_plus after saddle point method}), we used the saddle point method. Thus, $u^{*}$ and $v_{*}$ are given by
\begin{eqnarray}
(u^{*}, v^{*})
= \mathrm{Extr}_{u \in [0,1],\, v \in [0,1]}
\{ \alpha - I(u,v) + F_{t}\varepsilon(u,v) \}\label{saddle criterion},
\end{eqnarray}
where $\mathrm{Extr}_{x} f(x)$ denotes the extremum of the function $f(x)$with respect to the variable $x$. 

Then, we determine these saddle points explicitly.
The rate function is calculated as follows:
\begin{eqnarray}
    I(u,v) &=& -\frac{1}{N}\log P (u,v | N_{+}^{t})\\
    &=&
    \frac{1+m\theta^{t}}{2}D_{\mathrm{KL}}\left(u \,\middle\|\, \frac{1+m}{2}\right)
    +
    \frac{1-m\theta^{t}}{2}D_{\mathrm{KL}}\left(v \,\middle\|\, \frac{1-m}{2}\right)
    \label{eq rate function},
\end{eqnarray}
where $D_{\mathrm{KL}}(x||y)$ represents the Kullback-Leibler divergence between two binary distributions: $D_{\mathrm{KL}}(x||y) = x\log\frac{x}{y} + (1-x)\log\frac{1-x}{1-y}$. To derive Eq. (\ref{eq rate function}), we used the formula of $P(u,v|N_{+}^{t})$ given as Eq. (\ref{eq P_uv}) and the Stirling formula. Then, using Eqs (\ref{eq rate function}), (\ref{eq constraint varepsilon}), and (\ref{saddle criterion}), $u_{t}^{*} := u^{*}$ and $v_{t}^{*} := v^{*}$ are obtained as
\begin{eqnarray}
    u_{t}^{*} = \frac{(1+m)(1+\theta^{t})}{2(1 + m\theta^{t})}\label{eq saddle u},\\
    v_{t}^{*} = \frac{(1-m)(1+\theta^{t})}{2(1 - m\theta^{t})}\label{eq saddle v}.
\end{eqnarray}
Substituting Eq. (\ref{eq saddle u}) and (\ref{eq saddle v}) into Eq. (\ref{eq def Z_plus after saddle point method}), we obtain
\begin{eqnarray}
    Z_{2...p_{+}} = e^{N[s_{t} + F_{t}\theta^{t}]},
\end{eqnarray}
where the typical value of the microcanonical entropy density $s_{t} := s(u^{*}, v^{*})$ is as follows:
\begin{eqnarray}
    \hspace{-15mm}s_{t} =
\alpha - \frac{1 + m}{2}
D_{\mathrm{KL}}\!\left(
\frac{1 + \theta^{t}}{2}
\,\middle\|\,
\frac{1 + m\theta^{t}}{2}
\right)
- \frac{1 - m}{2}
D_{\mathrm{KL}}\!\left(
\frac{1 + \theta^{t}}{2}
\,\middle\|\,
\frac{1 - m\theta^{t}}{2}
\right)
\label{eq st}.
\end{eqnarray}

As mentioned above, Eq. (\ref{eq st}) provides a simple explanation for why considering either $Z_{+}$ or $Z_{-}$ does not lead to any loss of generality.
In the case of $Z_{-}$, the data distribution of single degree of freedom is $P_{0i}^{-}(x_{i}^{\mu})$, where $P_{0i}^{-}$ denotes the single-site factor of $P^-$, given in Eq. (\ref{eq def P_plus and P_minus}). Hence, the transformation $m \rightarrow -m$ should be made.  Similarly, for $Z_{-}$ we have $N_{+}^{t} = \frac{1 - m\theta^{t}}{2}$, hence the transformation $m\theta^{t} \rightarrow -m\theta^{t}$ should be made.
Since the Kullback–Leibler divergence in Eq. (\ref{eq st}) is clearly invariant under these two transformations, Eq. (\ref{eq st}) also holds for the case $Z_{-}$.

\refstepcounter{section}
\section*{Appendix \thesection. The derivation of the cloning probability for the empirical data}
\addcontentsline{toc}{section}{Appendix \thesection. The derivation of the cloning probability for the empirical data}
\label{app:appB}

\renewcommand{\theequation}{\thesection.\arabic{equation}}
\setcounter{equation}{0}

\refstepcounter{subsection}
\subsection*{\thesubsection\ Cloning for speciation time}
\label{app:appB_1}

We describe the evaluation of the cloning probability based on the empirical data, which we use to validate $t_{S}$ and $t_{C}$ in the real-data experiments. We first explain the evaluation of $\phi^{e}_{S}(t)$, defined as the probability that two samples sharing the same configuration at a given time belong to the same class at time $0$. Each data point in the empirical dataset $\mathcal{D} = \{ \bm x^{\mu} \}_{\mu=1}^{p}$ belongs to either class $C_{1}$ or class $C_{2}$, with fraction $\eta$ for $C_{1}$ and $1-\eta$ for $C_{2}$, where $0<\eta<1$. The probability that the system takes the configuration $\bm x_{t}$ at time $t$, conditioned on belonging to class $C_{1}$ at time $0$, can be written in terms of the probability distribution function $P(\bm x_{t} |\bm x_{0})$ (Eq. (\ref{eq forward 0tot})) as follows.
\begin{eqnarray}
    P(\bm x_{t} | C_{1}) = \sum_{\mu \in C_{1}} \frac{e^{F_{t} \bm x_{t} \cdot \bm x^{\mu}}}{[2\cosh F_{t}]^{N}}.
\end{eqnarray}
We here used the relation: $(1 + \tanh(F)S)/2 = e^{FS}/2\cosh F_{t}$, where $F \in \mathbb{R}^{1}, S = \{-1,1\}$, for the form of $P(\bm x_{t} |\bm x_{0})$ given by Eq. (\ref{eq forward 0tot}). The joint probability that the system takes the configuration $\bm x_{t}$ at time $t$ and the data $\bm x^{\mu}$ is sampled from $C_{1}$ at $t = 0$ becomes
\begin{eqnarray}
    P(\bm x_{t}, C_{1}) = \eta \sum_{\mu \in C_{1}} \frac{e^{F_{t} \bm x_{t} \cdot \bm x^{\mu}}}{[2\cosh F_{t}]^{N}}.
\end{eqnarray}
The probability of belonging to class $C_{1}$ at time $t = 0$, conditioned on observing $\bm x_{t}$ at time $t$, is calculated.
\begin{eqnarray}
    P(C_{1} | \bm x_{t}) &=& \frac{P(\bm x_{t}, C_{1})}{P(\bm x_{t})}\\
    &=&
    \frac{P(\bm x_{t}, C_{1})}{P_{t}(\bm x_{t})}\\
    &=&
    \frac{\eta \sum_{\mu \in C_{1}} \frac{e^{F_{t} \bm x_{t} \cdot \bm x^{\mu}}}{[2\cosh F_{t}]^{N}}}{\eta\prod_{i = 1}^{N}\frac{1 + \theta^{t} m x_{ti}}{2} + (1-\eta) \prod_{i = 1}^{N}\frac{1 - \theta^{t} m x_{ti}}{2}}\\
    &=&
    \frac{\eta \sum_{\mu \in C_{1}} \frac{e^{F_{t} \bm x_{t} \cdot \bm x^{\mu}}}{[2\cosh F_{t}]^{N}}}{\eta \sum_{\mu \in C_{1}} \frac{e^{F_{t} \bm x_{t} \cdot \bm x^{\mu}}}{[2\cosh F_{t}]^{N}} + (1-\eta) \sum_{\mu \in C_{1}} \frac{e^{F_{t} \bm x_{t} \cdot \bm x^{\mu}}}{[2\cosh F_{t}]^{N}}}\\
    &=&
    \frac{\eta \sum_{\mu \in C_{1}} e^{F_{t} \bm x_{t} \cdot \bm x^{\mu}}}
    {
    \eta \sum_{\mu \in C_{1}}
            e^{F_{t} \bm x_{t} \cdot \bm x^{\mu}}
             + (1-\eta) \sum_{\mu \in C_{1}}
            e^{F_{t} \bm x_{t} \cdot \bm x^{\mu}}
             }\label{eq result of P_C1 given xt}.
\end{eqnarray}

Next, we derive the probability that two clones sharing the configuration $\bm x_{t}$ at time $t$ belong to the same class at time $0$, and denote it by $R_{S}^{e}(\bm x_{t})$. Using the fact that the probability that both trajectories belong to a given class $C$ at time $0$ is $P(C | \bm x_{t})^{2}$, and using Eq. (\ref{eq result of P_C1 given xt}), $R_{S}^{e}(\bm x_{t})$ can be written as follows.
\begin{eqnarray}
    R_{S}^{e}(\bm x_{t}) = \frac{\left(\eta\sum_{\mu \in C_{1}} e^{F_{t} \bm x_{t} \cdot \bm x^{\mu}}\right)^{2} + \left((1-\eta)\sum_{\mu \in C_{2}} e^{F_{t} \bm x_{t} \cdot \bm x^{\mu}}\right)^{2}}{
    \left(\eta\sum_{\mu \in C_{1}} e^{F_{t} \bm x_{t} \cdot \bm x^{\mu}} + (1-\eta)\sum_{\mu \in C_{2}} e^{F_{t} \bm x_{t} \cdot \bm x^{\mu}}\right)^{2}
    }.
\end{eqnarray}
The cloning probability for speciation using empirical data, $\phi_{S}^{e}(t)$, is given by following expectaion:
\begin{eqnarray}
    \phi_{S}^{e}(t) = \sum_{\bm x_{t}} R_{S}^{e}(\bm x_{t}) P_{t}^{e}(\bm x_{t})\\
     =
    \sum_{\bm x_{t}} \frac{\left(\eta\sum_{\mu \in C_{1}} e^{F_{t} \bm x_{t} \cdot \bm x^{\mu}}\right)^{2} + \left((1-\eta)\sum_{\mu \in C_{2}} e^{F_{t} \bm x_{t} \cdot \bm x^{\mu}}\right)^{2}}{
    \left(\eta\sum_{\mu \in C_{1}} e^{F_{t} \bm x_{t} \cdot \bm x^{\mu}} + (1-\eta)\sum_{\mu \in C_{2}} e^{F_{t} \bm x_{t} \cdot \bm x^{\mu}}\right)^{2}
    }\nonumber\\
    \hspace{20mm} \times \frac{1}{p} \frac{1}{[2\cosh F_{t}]^{N}} \left( \sum_{\mu = 1}^{p} e^{F_{t} \bm x_{t}\cdot \bm x^{\mu}}\right)\label{eq phi_S_e}.
\end{eqnarray}
An analytic calculation or a direct computation of Eq. (\ref{eq phi_S_e}) is difficult. Therefore, we evaluate $\phi_{S}^{e}(t)$ as following empirical average:
\begin{eqnarray}
    \phi_{S}^{e}(t)\approx \frac{1}{\omega} \sum_{\nu = 1}^{\omega}R_{S}^{e}(\bm x_{t}^{(\nu)}) P_{t}^{e}(\bm x_{t}^{(\nu)})\label{eq phi_S_e empirical ave },
\end{eqnarray}
where $\omega$ is the sample size for this this sampling. 

\refstepcounter{subsection}
\subsection*{\thesubsection\ Cloning for collapse time}

The cloning probability for the validation of collapse time based on the empirical data can be similarly understood by the explanation in Sec. \ref{app:appB_1}. Here, the cloning probability refers to the probability that two clones reach the \textit{same data point} at time $t = 0$. Therefore, by replacing the classes with individual data points in the discussion presented in \ref{app:appB_1} above, the desired cloning probability can be derived in exactly the same manner.

Suppose that the $\mu$-th data point $\bm x^{\mu}$ is sampled at time $0$. Under this condition, the probability that the system takes the value $\bm x_{t}$ at time $t$ in the forward process is given by
\begin{eqnarray}
    P(\bm x_{t} |\bm x_{0} =  \bm x^{\mu}) = \frac{e^{F_{t} \bm x_{t} \cdot \bm x^{\mu}}}{[2\cosh F_{t}]^{N}}.
\end{eqnarray}
The joint probability that the system takes the configuration $\bm x_{t}$ at time $t$ and the data $\bm x^{\mu}$ is sampled at $t = 0$ is
\begin{eqnarray}
    P(\bm x_{t}, \bm x_{0} = \bm x^{\mu}) = \frac{1}{p}\frac{e^{F_{t} \bm x_{t} \cdot \bm x^{\mu}}}{[2\cosh F_{t}]^{N}}.
\end{eqnarray}
Thus, the probability of appearance $\bm x^{\mu}$ at $t = 0$, conditioned on observing $\bm x_{t}$ at time $t$ becomes as follows:
\begin{eqnarray}
    P(\bm x_{0} = \bm x^{\mu} | \bm x_{t}) & = & \frac{P(\bm x_{t}, \bm x_{0} = \bm x^{\mu})}{P_{t}(\bm x_{t})}\\
    &=&
    \frac{e^{F_{t} \bm x_{t} \cdot \bm x_{\mu}}}{\sum_{\nu = 1}^{p}e^{F_{t} \bm x_{t} \cdot \bm x_{\nu}}}\label{eq B15}.
\end{eqnarray}
Eq. (\ref{eq B15}) can be derived by following the same discussion described in Sec. \ref{subsec:sec5_1} and Sec. \ref{app:appB_1}. The probability that two clones sharing the configuration $\bm x_{t}$ at time $t$ reach the same data point $\bm x^{\mu^{\prime}}$ is $\left( P(\bm x_{0} = \bm x^{\mu^{\prime}} | \bm x_{t})\right)^{2}$. Therefore, the probability that two clones $\bm x_{t}$ reach the same arbitrary data point at time $t = 0$ is given by
\begin{eqnarray}
    R_{C}^{e}(\bm x_{t}) &=& \sum_{\mu = 1}^{p} \left( P(\bm x_{0} = \bm x^{\mu} | \bm x_{t})\right)^{2}\\
    &=&
    \sum_{\mu = 1}^{p} \left(  \frac{e^{F_{t} \bm x_{t} \cdot \bm x_{\mu}}}{\sum_{\nu = 1}^{p}e^{F_{t} \bm x_{t} \cdot \bm x_{\nu}}} \right)^{2}.
\end{eqnarray}
Then, the cloning probability for collapse using empirical data, $\phi_{C}^{e}(t)$, is given by following expectaion:
\begin{eqnarray}
    \phi_{C}^{e}(t) &=& \sum_{\bm x_{t}} R_{C}^{e}(\bm x_{t}) P_{t}^{e}(\bm x_{t})\\
    &=&
    \sum_{\bm x_{t}} 
    \sum_{\mu = 1}^{p} \left(  \frac{e^{F_{t} \bm x_{t} \cdot \bm x^{\mu}}}{\sum_{\nu = 1}^{p}e^{F_{t} \bm x_{t} \cdot \bm x^{\nu}}} \right)^{2} \frac{1}{p} \sum_{\sigma = 1}^{p}\frac{e^{F_{t} \bm x_{t} \cdot \bm x^{\sigma}}}{[2\cosh F_{t}]^{N}}\label{eq phi_C_t}.
\end{eqnarray}
As in the case of $\phi_{S}^{e}(t)$, we evaluate the expectation Eq. (\ref{eq phi_C_t}) by following empirical average:
\begin{eqnarray}
    \phi_{C}^{e}(t) \approx \frac{1}{\pi} \sum_{\nu^{\prime} = 1}^{\pi}R_{C}^{e}(\bm x_{t}^{(\nu^{\prime})}) P_{t}^{e}(\bm x_{t}^{(\nu^{\prime})})\label{eq phi_C_e empirical ave},
\end{eqnarray}
where $\pi$ is the sample size for this sampling. 

\refstepcounter{section}
\section*{Appendix \thesection. The learning in discrete diffusion models}
\addcontentsline{toc}{section}{Appendix \thesection. The learning in discrete diffusion models}
\label{app:appC}

\renewcommand{\theequation}{\thesection.\arabic{equation}}
\setcounter{equation}{0}

The notation used in the following description has already been defined in Sec. \ref{sec:sec2} of the main text. In particular, $\bm z_t$ denotes the one-hot vector representing the data at time $t$, and this representation is common to all $i = 1,2,\cdots,N$. The following explanation is mainly based on \cite{hoogeboom2022autoregressive}.

The backward transition probability $p_{\Theta}(\bm z_{t-1} | \bm z_t)$ is modeled as a categorical distribution parameterized by a neural network.
Given the input $(\bm z_t,t)$, the neural network outputs a vector of logits
\begin{eqnarray}
\bm h_{\Theta}(\bm z_t,t) = 
(h_{\Theta,1}(\bm z_t,t),\cdots,h_{\Theta,K}(\bm z_t,t)).
\end{eqnarray}
These logits are transformed into a probability vector through the softmax function,
\begin{eqnarray}
\pi_{\Theta,k}(\bm z_t,t) = 
\frac{\exp\left(h_{\Theta,k}(\bm z_t,t)\right)}
{\sum_{j=1}^{K}\exp\left(h_{\Theta,j}(\bm z_t,t)\right)},
\qquad k=1,\cdots,K .
\end{eqnarray}
The resulting probability vector
\begin{eqnarray}
\bm \pi_{\Theta}(\bm z_t,t) = 
(\pi_{\Theta,1}(\bm z_t,t),\cdots,\pi_{\Theta,K}(\bm z_t,t))
\end{eqnarray}
defines the categorical distribution governing the backward process.

Since the state $\bm z_{t-1}$ is represented as a one-hot vector, the log-likelihood of the backward transition takes the form
\begin{eqnarray}
\log p_{\Theta}(\bm z_{t-1} | \bm z_t) = 
\sum_{k=1}^{K}
z_{t-1,k}
\log
\pi_{\Theta,k}(\bm z_t,t).
\end{eqnarray}
The parameters of the neural network are learned by maximizing the likelihood of the true state generated by the forward diffusion process.
Equivalently, the learning objective is given by minimizing the expectation of the negative log-likelihood,
\begin{eqnarray}
\mathcal{L} = 
\mathbb{E}_{q(\bm z_0)q(\bm z_t | \bm z_0)}
\left[
\log
p_{\Theta}(\bm z_{t-1} | \bm z_t)
\right],
\end{eqnarray}
where $q(\bm z_t | \bm z_0)$ denotes the forward diffusion process and $q(\bm z_0)$ is the data distribution.
In practice, since the data distribution $q(\bm z_0)$ is unknown, this expectation is approximated by the empirical average over the training dataset. The resulting objective corresponds to minimizing the cross-entropy between the predicted categorical distribution and the true one-hot state generated by the forward process.

\refstepcounter{section}
\section*{Appendix \thesection. Details of the real data experiments for the speciation}
\addcontentsline{toc}{section}{Appendix \thesection. Details of the real data experiments for the speciation}
\label{app:appE}

\renewcommand{\theequation}{\thesection.\arabic{equation}}
\setcounter{equation}{0}

In Sec. \ref{subsec:sec6_1} of the main text, we conducted empirical speciation experiments by training an actual discrete diffusion model on real-world datasets (binarized MNIST). Accordingly, this Appendix provides detailed descriptions of the model architecture, the training procedure, and representative generation results.
In contrast, the collapse experiments presented in Sec. \ref{subsec:sec6_2} rely solely on real data and do not involve training a discrete diffusion model. Since the dataset preparation for the collapse experiments is sufficiently described in Sec. \ref{subsec:sec6_2}, we focus here exclusively on the technical details relevant to the speciation experiments.

\refstepcounter{subsection}
\subsection*{\thesubsection\ Discrete Denoising Diffusion Probabilistic Models}

The Discrete Denoising Diffusion Probabilistic Models \cite{austin2021structured} (D3PMs) employed in this study constitute one of the most widely used and conceptually simplest classes of discrete diffusion models. Let $\bm z_{t}$ denotes, following the explanation of Sec. \ref{sec:sec2}, one-hot vector with respect to the categories irrespective the index $i$. The loss function of the D3PM is given by
\begin{eqnarray}
    L_{\lambda} = L_{\mathrm{vb}} + \lambda \mathbb{E}_{q_{0}(\bm z_{0})} \mathbb{E}_{q(\bm z_{t} | \bm z_{0})} \left[ -\log \tilde{p}_{\Theta}(\bm z_{0} | \bm z_{t})\right]\label{eq def d3pm loss},
\end{eqnarray}
where $L_{\mathrm{vb}}$ denotes the loss function of the DDPM \cite{ho2020denoising} given by
\begin{eqnarray}
L_{\mathrm{vb}} 
&= \mathbb{E}_{q(\bm x_{1:T} | \bm x_{0})} \Big[
D_{\mathrm{KL}}(q(\bm x_{T} | \bm x_{0}) \,\|\, p_{T}(\bm x_{T})) \nonumber \\
&\quad + \sum_{t = 2}^{T} 
D_{\mathrm{KL}}( q(\bm x_{t-1}|\bm x_{t}, \bm x_{0}) 
\,\|\, p_{\Theta}(\bm x_{t-1} | \bm x_{t}) ) - \log p_{\Theta}(\bm x_{0} | \bm x_{1})
\Big]\label{eq def ddpm loss}.
\end{eqnarray}
The posterior $q(\bm z_{t-1} | \bm z_{t}, \bm z_{0})$ can be analytically obtained as a Categorical distribution, just as it is Gaussian distribution in the continuous-data case (DDPM). In Eq. (\ref{eq def d3pm loss}), the likelihood $\tilde{p}_{\Theta}(\bm z_{0} | \bm z_{t})$ is a learnable categorical distribution parameterized by a neural network. The second term of the right hand side of Eq. (\ref{eq def d3pm loss}) is added for a stronger supervision to the data $\bm x_{0}$\cite{austin2021structured}. The symbol $\lambda$ denotes the control parameter for the second term. At the backward process after the optimization $\Theta$ in the loss function, the transition probability is calculated as follows:
\begin{eqnarray}
    p_{\Theta}(\bm z_{t-1} | \bm z_{t}) = \sum_{\tilde{\bm z}_{0}} q(\bm z_{t-1} | \bm z_{t}, \tilde{\bm z}_{0}) \tilde{p}_{\Theta}(\tilde{\bm z}_{0} | \bm z_{t}),
\end{eqnarray}
where, $\tilde{\bm z}_{0}$ denotes a dummy variable introduced 
to distinguish it from the training data $\bm z_{0}$.
In practice, both $\tilde{p}_{\Theta}(\tilde{\bm z}_{0} | \bm z_{t})$ and 
$p_{\Theta}(\bm z_{t-1} |\bm z_{t})$ are computed by estimating their logits 
with a neural network and applying the softmax function to obtain categorical distributions.

\refstepcounter{subsection}
\subsection*{\thesubsection\ Details of the Learning setup}

BinMNIST consists of binary-valued pixels taking values in $\{0,1\}$, where 0 represents black and 1 represents white. The representation $\{-1,1\}$ is used only at the generation stage shown in Fig. 4. The remaining dataset details are summarized in Sec. \ref{subsec:sec6_1} of the main text. 

The settings of the discrete diffusion model are as follows. The number of diffusion timesteps is $T =500$. The transition probability matrix between categories is the uniform transition matrix, as described in Sec. \ref{sec:sec2} of the main text. The noise schedule is linear, as stated in Sec. \ref{subsec:sec6_1}. The noise strength at each timestep $\beta_{t}$ is linearly interpolated from $10^{-4}$ to $0.02$. Therefore, the slope $a$ and intercept $b$ used in the main text are respectively $a = (0.02 - 10^{-4}) / (500 - 1)$ and $b = 10^{-4}$. We set $\lambda = 0.05$. Class-conditional generation is performed. 

The details of the neural network training is as follows. The neural network used is a lightweight UNet without attention layers. The optimizer is Adam. The learning rate is set to $10^{-5}$ and the batch size is 128. The total number of training steps is 150,000.

\refstepcounter{subsection}
\subsection*{\thesubsection\ The generated images of the binarized MNIST}

The generated images of BinMNIST corresponding to labels 1 and 8 are shown in Fig. 7, including those at every 100 timesteps as well as at $t = t_{S}^{\mathrm{lin}}$. Since images are generated only at integer timesteps, the sample displayed at $t = t_{S}^{\mathrm{lin}}$ corresponds to $t = 293$, given that the speciation time is estimated as $t_{S}^{\mathrm{lin}} = 293.80$ according to Sec. \ref{subsec:sec6_1} of the main text.

From this figure, it is difficult to clearly discern the degree of commitment to each label at $t = t_{S}^{\mathrm{lin}}$. However, at the immediately preceding timestep, $t = 200$, a slight branching toward the characteristic shape of each digit can already be observed, indicating that the generated images indeed begin to exhibit commitment to their respective labels.

As mentioned above, we do not perform training or generation on the MovieLens Tag Genome dataset; therefore, no results are shown. 

\begin{figure}[h]
\centering
 \includegraphics[width=9cm]{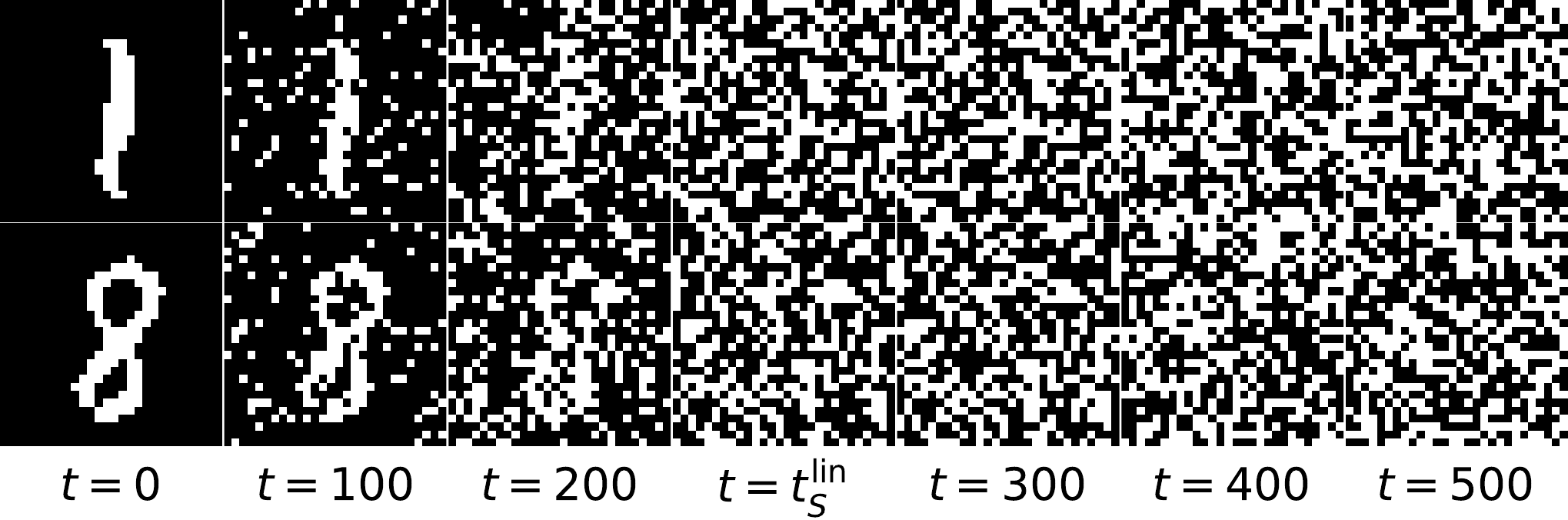}
 \caption{The generated images of BinMNIST of the label 1 and 8 at every 100 timesteps as well as at $t = t_{S}^{\mathrm{lin}}$. Since images are generated only at integer timesteps, the sample displayed at $t = t_{S}^{\mathrm{lin}}$ corresponds to $t = 294$, given that the speciation time is estimated as $t_{S}^{\mathrm{lin}} = 293.80$ according to Sec. \ref{subsec:sec6_1} of the main text. For each label, we select from the 10 generated image sequences the one whose overlap with the corresponding training-data mean vector $\bar{\bm x}^{(1)}$ or $\bar{\bm x}^{(8)}$ at $t = 0$ is maximal. }
\end{figure}

\addcontentsline{toc}{section}{References}
\section*{References}
\bibliography{Dynamical_regimes_of_discrete_diffusion_models}
\bibliographystyle{unsrt}

%\maketitle

\end{document}